\documentclass{jfm}
\usepackage{graphicx}
\usepackage{epstopdf, epsfig}
\usepackage{amsmath}
\usepackage{xcolor}
\usepackage{hyperref}
\hypersetup{hidelinks,
	colorlinks=true,
	allcolors=blue,
	citecolor=blue,
	citebordercolor=blue,
	pdfstartview=Fit,
	breaklinks=true,
	linkbordercolor=blue,
}
\usepackage{soul}
\soulregister\cite7
\soulregister\citep7
\soulregister\citet7
\soulregister\ref7

\shorttitle{Hydrodynamic coupling of a cilia-mucus system}
\shortauthor{Q. Mao, U. D'Ortona and J. Favier}

\title{Hydrodynamic coupling of a cilia-mucus system in Herschel-Bulkley flows}

\author{Q. Mao\aff{1},
        U. D'Ortona\aff{1}
   \and J. Favier\aff{1}
   \corresp{\email{julien.favier@univ-amu.fr}}}

\affiliation{\aff{1}Aix-Marseille Univ., CNRS, Centrale Marseille, M2P2, Marseille, France}

\begin{document}

\maketitle

\begin{abstract}
The yield stress and shear thinning properties of mucus are identified as critical for the ciliary coordination and the mucus transport in the human airways. We use here numerical simulations to explore the hydrodynamic coupling of cilia and mucus with these two properties using the Herschel-Bulkley model, in a lattice Boltzmann solver for the fluid flow. Three mucus flow regimes, i.e. a poorly organized regime, a swirly regime, and a fully unidirectional regime are observed and analysed by parametric studies. We systematically investigate the effects of ciliary density, interaction length, Bingham number and flow index on the mucus flow regime formation. The underlying mechanism of the regime formation is analysed in detail by examining the variation of two physical quantities (polarization and integral length) and the evolution of the flow velocity, viscosity, and shear rate fields. Mucus viscosity is found to be the dominant parameter influencing the regime formation when enhancing the yield stress and shear thinning properties. The present model is able to reproduce the solid body rotation observed in experiments \citep{LoiseauGsell20}. A more precise prediction can be achieved by incorporating non-Newtonian properties into the modeling of mucus as proposed by \cite{GsellLoiseau20}.
\end{abstract}


\section{Introduction}

\noindent Mucociliary clearance driven by ciliary beating in the human airways has received much attention due to its critical role in the capture and clearance of foreign pollutants and pathogens \citep{WannerSalathe96,Grotberg21,SedaghatBehnia23}. The human airways are protected by two fluid layers, a periciliary layer (PCL) covering the epithelial surface and a mucus layer on top of the PCL \citep{ChilversOCallaghan00,KnowlesBoucher02,ChoudhuryFiloche23}. The mucus is often described as a yield stress and shear thinning fluid \citep{BanerjeeBellare01,NordgardDraget11,ChatelinAnne-Archard17}. Cilia are almost immersed in the PCL and interact with the mucus through their tips. The force generated by the cilia propels the mucus flow and the mucus in turn affects the orientation of the ciliary beating \citep{LoiseauGsell20,GsellLoiseau20,PellicciottaHamilton20}. This hydrodynamic coupling between mucus and millions of microscopic cilia has a major impact on ciliary coordination (self-organization) and mucus transport. Understanding the hydrodynamic mechanism of cilia-mucus interaction is desirable for the study of various respiratory diseases caused by the impairment of mucus transport.

Considerable research effort has been devoted to the ciliary coordination and cilia-induced flow. For the former, the proposal of a cilia model can be traced back to an analytical study by \citet{BartonRaynor67}, where the cilium was simplified as an oscillating cylinder mounted on a plate. The relationship between the flow rate and the geometric parameters of the cilium was determined. Since then, many other studies involving modeling flexible cilia have been reported. Two asymmetric beating phases of a cilium were identified, i.e. an effective stroke characterized by almost straight cilium to better drive the mucus and a recovery stroke characterized by large deformed cilium to reduce the retarding effect on the mucus \citep{Blake72,XuJiang19}. Cilia in an array can coordinate with each other to generate metachronal waves instead of repeating two beating phases synchronously \citep{HussongBreugem11,ElgetiGompper13,MengBennett21,MesdjianWang22,WangTang22}. This usually depends on the phase difference between adjacent cilia \citep{ChateauFavier17,HallClarke20}, and is related to ciliary flexibility \citep{KimNetz06} and ciliary density \citep{ChateauDOrtona18}. Various metachronal waves have been observed, such as the antipleptic and symplectic waves. These two waves move in the opposite and the same direction of the flow, respectively. Cilia beating with an antiplectic wave were found to be more efficient in transporting and mixing fluid than cilia beating synchronously or with a symplectic wave. On the other hand, various experimental and numerical studies have investigated the cilia-induced flow, usually focusing on local flow characteristics and flow rate \citep{BrumleyWan14,WeiDehnavi19,WeiDehnavi21,BoselliJullien21,HuMeng23}. Among them, \citet{BrumleyWan14} measured the flow around a single cilium and a pair of cilia, and calculated the instantaneous forces generated by the cilia using a Stokeslet model. They highlighted the importance of hydrodynamic coupling; a synchronized beating of two cilia can be realized even when only hydrodynamic interactions exist. Furthermore, \citet{DingNawroth14} observed a transport region and a mixing (shear) region above and below the ciliary tips, respectively. The asymmetric stroke of the cilia and the no-slip epithelial surface resulted in a shear-like flow field. The enhancement of fluid transport and mixing was mainly attributed to the increase in the shear rate. Fluid transport was also found to be continuous even though the epithelial surface was not completely covered by the cilia \citep{JuanMathijssen20}.

A long-range ciliary coordination distinct from metachronal waves has been discovered \citep{MatsuiGrubb98,Tarran05,ShapiroFernandez14,KhelloufiLoiseau18}, characterized by large-scale mucus swirls accompanied by cilia beating in a circular pattern. Several experimental studies have shown the existence of hydrodynamic coupling between the ciliary coordination and the circular mucus flow \citep{MitchellJacobs07,Guirao10,FaubelWestendorf16}. Recently, \citet{LoiseauGsell20} and \citet{GsellLoiseau20} experimentally investigated the hydrodynamic coupling of the cilia-mucus system in detail and proposed a two-dimensional model to predict the ciliary coordination and the Newtonian mucus flow. They demonstrated that the hydrodynamic coupling of cilia and mucus dominates the long-range coordination. The formation of mucus swirls was closely related to the density and the interaction length of the cilia. As mentioned above, mucus is a non-Newtonian fluid with yield stress and shear thinning properties. Some researchers have investigated the effect of non-Newtonian properties on mucus transport \citep{ChatelinPoncet16,SedaghatGeorge21,SedaghatFarnoud22,Modaresi23,WangGsell23}. The present study aims to focus first on the yield stress and shear thinning properties although the mucus rheology also exhibits other properties, e.g. viscoelasticity \citep{Vasquez2016modeling,Guo2017computational,ChoudhuryFiloche23}. Most of the studies employed numerical methods because the non-Newtonian properties can be well controlled. However, few studies have been reported on the effect of non-Newtonian properties on long-range ciliary coordination, which warrants a more detailed investigation.

The objective of the present study is to numerically explore the hydrodynamic coupling of cilia and mucus with yield stress and shear thinning properties. The mucus flow is solved by the Lattice-Boltzmann (LB) method. The cilia-mucus interaction is handled by an alignment rule and the non-Newtonian fluid is modelled by the Herschel-Bulkley model. The effects of ciliary density ($\phi$), interaction length ($\lambda$), Bingham number ($Bn$, quantifying yield stress effect) and flow index ($n$, quantifying shear thinning effect) on the formation of mucus flow (or ciliary beating orientation) regime are examined. Three different mucus flow regimes are observed: a poorly organized (PO) regime, a swirly (S) regime, and a fully unidirectional (FU) regime (corresponding to the poorly aligned, swirly and fully aligned regimes in \citet{GsellLoiseau20}). Two physical parameters, i.e. polarization ($P$) and integral length ($\Lambda$), are used to identify the three regimes. The mechanism of regime formation caused by the yield stress and shear thinning effects is characterized by the evolution of the flow velocity, viscosity, and shear rate fields. In addition, a rescaling of $\lambda$ is proposed for different $Bn$ and $n$.

\section{Computational model}\label{sec:computational_model}

\begin{figure}
\centerline{\includegraphics[width=0.7\linewidth]{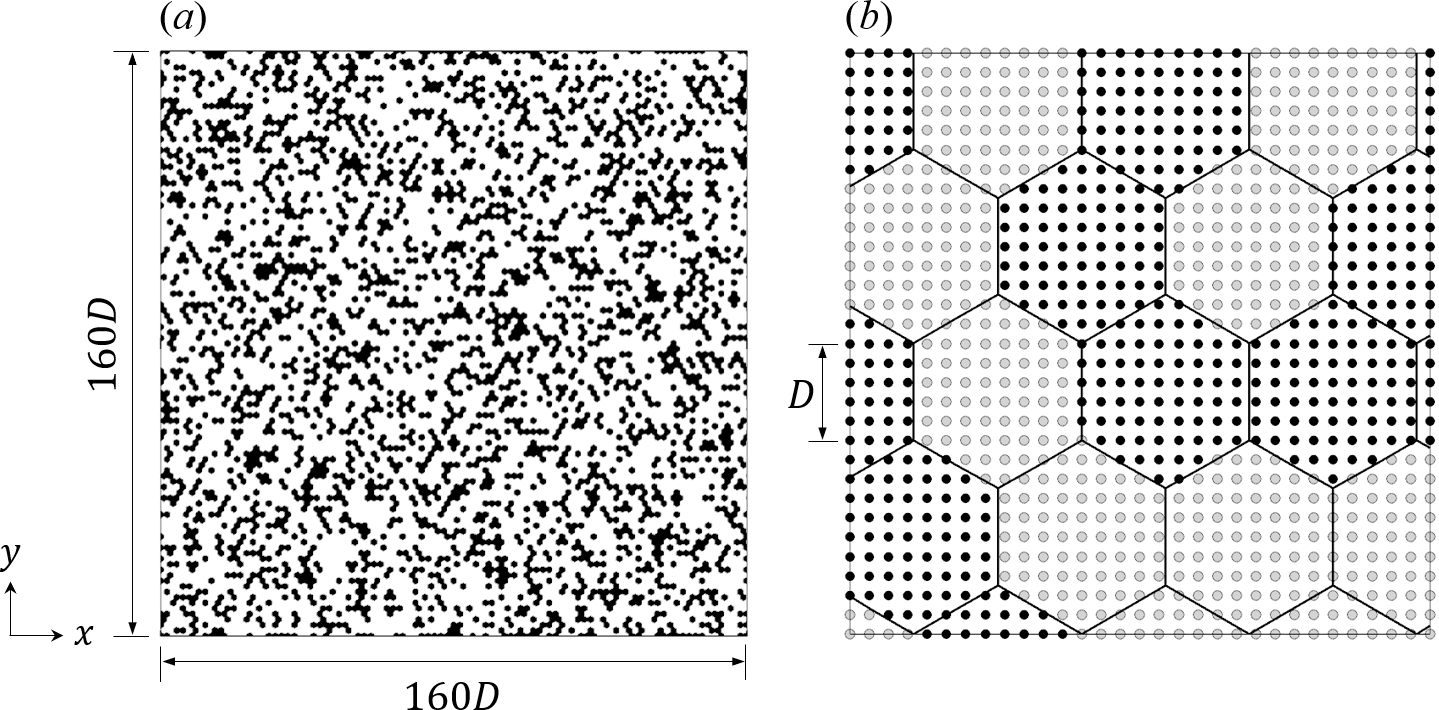}}
\caption{($a$) Visualization of the computational domain (ciliary density $\phi=0.3$). Cells are discretized using hexagonal elements, with ciliated elements (black) randomly placed during initialization. ($b$) Closer visualization of the hexagonal elements and the underlying lattice nodes. The black and grey dots represent ciliated and non-ciliated nodes, respectively.}
\label{figure01}
\end{figure}

\noindent Mucus flow is almost parallel to the epithelium and uniform along the direction perpendicular to the epithelium \citep{GsellLoiseau20}. In addition, we concentrate on long-range fluid flow parallel to the epithelium, whose length scales are much larger than the typical fluid layer thickness. Therefore, the mucus flow is considered to be two-dimensional and a 2D hydrodynamic model is sufficient to describe the mucus motion. The flow is predicted based on a lattice-Boltzmann (LB) solver and the interaction between the ciliary beating and the mucus motion is handled by an alignment rule. A visualization of the computational domain for $\phi=0.3$ is shown in figure \ref{figure01}($a$). The square domain is approximately 160$D$ in both length and height. The cells are discretized using hexagonal elements, where the ciliated elements (black) are randomly placed during initialization. A hexagonal element represents a patch containing several ciliated cells with a common direction of ciliary beating. About 10$^4$ elements are contained in the domain. Here, $D$ represents the side length of the hexagonal elements and $\phi=A_c/A$ represents the ciliary density, where $A_c$ is the ciliated area and $A$ is the total area. A closer visualization of the hexagonal elements and the underlying lattice nodes is shown in figure \ref{figure01}($b$). The computational domain is discretized on a uniform Cartesian grid. The black and grey dots represent ciliated and non-ciliated nodes, respectively. Note that cilia are simplified as ciliated nodes. Ciliary beating is modeled by a force constant in magnitude, whose orientation can change over time \citep{GsellLoiseau20}. The initial condition is $\boldsymbol{u}=0$ with random orientation of ciliary forces and random placement of ciliated elements. Periodic boundary conditions are specified at the domain boundaries.

In the LB method \citep{Kruger2017lattice,MaWang20,LuLei22}, the particle distribution function $f(\boldsymbol{x},\boldsymbol{\xi},t)$ is used to describe the mucus motion, representing the density of fluid particles moving with velocity $\boldsymbol{\xi}$ at location $\boldsymbol{x}$ and time $t$. The dynamics of $f(\boldsymbol{x},\boldsymbol{\xi},t)$ is governed by the Boltzmann equation:
\begin{equation}
\displaystyle
    \frac{\partial f}{\partial t} + \boldsymbol{\xi} \cdot  \nabla f= \Gamma (f),
    \label{eq_lbe}
\end{equation}
where $\Gamma$ is the collision operator. Equation (\ref{eq_lbe}) is equivalent to the Navier-Stokes equations at the macroscopic level \citep{Kruger2017lattice}. The lattice-Boltzmann equation is obtained by discretizing equation (\ref{eq_lbe}) in velocity space, physical space and time. A set of velocity vectors \{$\boldsymbol e_l$, $l=0$, ..., $Q-1$\} is used to discretize the velocity space, where $Q$ is the number of discrete velocities. The $D2Q9$ scheme is employed as the discretization model to discretize the velocity space by nine velocities:
\begin{center}
\begin{equation}
\displaystyle
\boldsymbol{e}_l=
\left\{
    \begin{array}{ll}
         (0,0),& l = 0, \\
         c\Big( \cos{(\frac{\pi(l-1)}{2})},\sin{(\frac{\pi(l-1)}{2})}\Big),& l\in[1,4], \\
         \sqrt{2}c\Big( \cos{(\frac{\pi(2l-9)}{4})},\sin{(\frac{\pi(2l-9)}{4})}\Big),& l\in[5,8],
    \end{array}{}
\right.
\end{equation}
\end{center}
where $c=\Delta x/\Delta t=\Delta y / \Delta t$ is the lattice velocity. As mentioned before, the computational domain is discretized on a uniform Cartesian grid, i.e. $\Delta h=\Delta x=\Delta y$ and $\Delta h=\Delta t=D/5=1$. The Lattice-Boltzmann equation is written as following, normalizing all the quantities by $c$ and $\Delta t$ and introducing an external body force:
\begin{equation}
\displaystyle
    f_l(\boldsymbol{x}+\boldsymbol{e}_l,t + 1) -f_l(\boldsymbol{x},t)= \Gamma_{l}(\boldsymbol{x},t) +  S^*_l(\boldsymbol{x},t),
    \label{discret_lbm_force}
\end{equation}
where $S^*_l(\boldsymbol{x},t)$ is the external body force term. The left and right sides of equation (\ref{discret_lbm_force}) are the streaming and collision steps respectively. These two steps can be treated separately due to the explicit equation (\ref{discret_lbm_force}). A two-relaxation-time collision operator is used in the present study:
\begin{equation}
    \Gamma_l = -\frac{1}{\tau^+}(f_l^+-f_l^{eq+})-\frac{1}{\tau^-} (f_l^- -f_l^{eq-}),
    \label{TRT_coll_step}
\end{equation}
where $\tau^+$ and $\tau^-$ are the symmetric and anti-symmetric relaxation times. $f_l^+$ and $f_l^-$ are the symmetric and anti-symmetric parts of $f_l$. The kinematic fluid viscosity $\nu=c_s^2(\tau^+-\frac{1}{2})$ is determined by $\tau^+$, where $c_s=1/\sqrt{3}$ is the lattice sound speed. $\tau^-$ is determined by the parameter $\Lambda_\tau=(\tau^+-0.5)(\tau^- -0.5)$. $\Lambda_\tau$ is kept constant to ensure the viscosity-independence \citep{GsellDOrtona21} and is set to $1/4$ according to the previous study \citep{GinzburgdHumieres10}. $f_l^{eq}$ is the equilibrium particle distribution function, expressed as:
\begin{equation}
\begin{array}{c}
\displaystyle
f_l^{eq}(\boldsymbol{x},t) = w_l\rho \Big[1+\frac{\boldsymbol{e}_l \cdot \boldsymbol{u}}{c_s^2}+\frac{(\boldsymbol{e}_l \cdot \boldsymbol{u})^2}{2c_s^4}-\frac{\boldsymbol{u}^2}{2c_s^2}\Big],
\end{array}{}
\label{pop_equilibre}
\end{equation}
where $w_l$ are the lattice weights and $\rho$ is the fluid density. In the present $D2Q9$ scheme $w_0=4/9$, $w_l=1/9$ for $l=1...4$ and $w_l=1/36$ for $l=5...8$ \citep{Qian1992lattice}. The symmetric and anti-symmetric parts of $f_l$ and $f_l^{eq}$ are expressed as:
\begin{equation}
\left\{   
    \begin{array}{cc}
         f_l^+=\dfrac{f_l+f_{\bar{l}}}{2}, & f_l^-=\dfrac{f_l-f_{\bar{l}}}{2},\\
         \\
         f_l^{eq+}=\dfrac{f_l^{eq}+f_{\bar{l}}^{eq}}{2}, & f_l^{eq-}=\dfrac{f_l^{eq}-f_{\bar{l}}^{eq}}{2},\\ 
   \end{array}{}
\right.
\label{trt_1}
\end{equation}
where the index $\bar l$ is defined such that $\boldsymbol{c_{\bar l}}=-\boldsymbol{c_l}$. The external body force term $S^*_l$ is expressed as:
\begin{equation}
\displaystyle
    S^*_l=\Big(1-\frac{1}{2\tau^+}\Big) S^+_l + \Big(1-\frac{1}{2\tau^-}\Big) S^-_l,
    \label{force_guo_trt}
\end{equation}
where $S^+_l=(S_l+S_{\bar{l}})/2$ and $S^-_l=(S_l-S_{\bar{l}})/2$ are the symmetric and anti-symmetric parts of $S_l$, which is expressed as:
\begin{equation}
\displaystyle
    S_l=w_l\Big[ \frac{\boldsymbol{e}_l-\boldsymbol{u}}{c_s^2}+ \frac{(\boldsymbol{e}_l \cdot \boldsymbol{u}) \cdot \boldsymbol{e}_l}{c_s^4}\Big] \boldsymbol{F}.
    \label{force_guo}
\end{equation}

The macroscopic quantities ($\rho$ and $\boldsymbol{u}$) are moments of the particle functions in the velocity space \citep{Kruger2017lattice}. $\rho$ is expressed as:
\begin{equation}
         \rho=\displaystyle \sum_{l=0}^8 f_l.
     \label{density}     
\end{equation}

The flow momentum corrected by the external forcing is:
\begin{equation}
    \displaystyle
         \rho\boldsymbol{u}=\displaystyle \sum_{l=0}^8 f_l \boldsymbol{e}_l + \dfrac{1}{2} \boldsymbol{F},\\
     \label{momentum}
\end{equation}
where the forcing $\boldsymbol{F}$ is the sum of the force $\boldsymbol{F}_c$ exerted by the cilia and the frictional force $\boldsymbol{F}_\nu$ generated by the PCL. $\boldsymbol{F}_c$ is imposed on the ciliated nodes only. In the present study, we are interested in the long-term dynamics of the flow, i.e. in time scales that are much larger than the ciliary beating period. Therefore, time-dependent beating is simplified to a point force. Its magnitude is the same for all ciliated nodes and is constant over time, while its orientation is determined by the alignment rule during the simulation. In addition, $\boldsymbol{F}_c$ is assumed to be independent of mucus properties, e.g. viscosity of the mucus. The orientation of $\boldsymbol{F}_c$ is the same for the ciliated nodes in the same hexagonal element. In the present study, the magnitude of $\boldsymbol{F}_c$ is set to be the same as that of $\boldsymbol{F}_\nu$ during the initialization. The frictional force is proportional to the fluid velocity ($\boldsymbol{F}_\nu=-\kappa \boldsymbol{u}$) and the PCL is assumed to be a Newtonian fluid, where $\kappa$ is the PCL friction coefficient. The frictional force is treated implicitly and equation (\ref{momentum}) becomes \citep{GsellLoiseau20}:
\begin{equation}
    \rho\boldsymbol{u}=\dfrac{\displaystyle \sum_{l=0}^8f_l\boldsymbol{e}_l+\frac{1}{2}\boldsymbol{F}_c}{1+\kappa/2\rho}.
    \label{momentum_cor} 
\end{equation}

Recall that the orientation $\theta_c^j$ of $\boldsymbol{F}_c$ on the $j$-th ciliated cell is determined by the alignment rule, which was inspired by certain experimental observations. The daily variations of the flow pattern of cilia-driven cerebrospinal fluid have been observed in in-vivo mouse brain ventricles \citep{FaubelWestendorf16}. In addition, \citet{Guirao10} showed that ciliary-beat orientations on cell cultures issued from the subventrical zone of newborn mice could be drastically changed by applying an external flow. The directional collective order of ciliary beats on the multiciliated skin cells of the Xenopus embryo could also be refined by applying an external flow to skin explants \citep{MitchellJacobs07}. Recently, it has been found that the directions of ciliary beating in the human airways tend to align progressively along mucus streamlines \citep{GsellLoiseau20,LoiseauGsell20}. A maximum angle reorientation of $35-40^{\circ}$ was shown. The direction of ciliary beating is stable over time and does not show reorientation when mucus was washed out. The reorientation of the cilia is not related to the motility of the cells because the tissue is jammed and the turnover of epithelial cells is very slow. These phenomena strongly suggest the existence of a coupling between hydrodynamics and long-range ciliary-beat orientation, inspiring the present alignment rule. $\Delta\theta=\theta_f^j-\theta_c^j$ represents the angle difference between the local flow ($\theta_f^j$) and the ciliary beating ($\theta_c^j$). The flow velocity is averaged over the $j$-th ciliated cell. The alignment rule is expressed as:
\begin{equation}
\displaystyle
\left\{
    \begin{array}{llll}
         \theta_c^j(t+\Delta t) & = &  \theta_c^j(t)+\Omega\dfrac{\Delta\theta^j(t)}{|\Delta\theta^j(t)|}\Delta t, \hspace{5mm} & \Delta\theta^j(t)>\theta_0,\\
         \theta_c^j(t+\Delta t) & = &  \theta_c^j(t), &\Delta\theta^j(t) \leq \theta_0,
    \end{array}
    \label{updatetheta}
\right.
\end{equation}
where $\Omega$ is a fixed angular velocity used to drive the reorientation of the ciliary beating, which does not affect the final steady solution \citep{GsellLoiseau20}. Its value is $\Omega=U_0/D$, where $U_0=0.01$ is the reference velocity (in lattice unit). $\theta_0$ is the angle threshold set to allow the steady solutions, i.e. $\theta_c^j(t+\Delta t)=\theta_c^j(t)$ when $\Delta\theta^j(t) \leq \theta_0$. The value of $\theta_0$ is very small ($\theta_0=2\Omega\Delta t=0.004$) to ensure a negligible influence on the final solutions. In summary, a two-way hydrodynamic coupling between the ciliary-beat orientation and the mucus motion is realized by the external forcing scheme and the alignment rule.

The Herschel-Bulkley model is employed to simulate non-Newtonian flows. The dynamic fluid viscosity $\mu$ is shear-dependent and is expressed as:
\begin{equation}
    \mu = \dfrac{\sigma_0}{\dot \gamma} + K \dot \gamma^{n-1},
    \label{HB}
\end{equation}
where $\sigma_0$ is the yield stress, $K$ is the flow consistency, $n$ is the flow index. When $n<1$, the viscosity decreases with increasing shear rate (shear thinning behavior). When $n>1$, the viscosity increases with increasing shear rate (shear thickening behavior). In the present study, only the shear thinning behavior is investigated. $\dot\gamma$ is the local shear-rate magnitude, which is expressed as:
\begin{equation}
    \dot\gamma=\sqrt{2(S_{11}^2+2S_{12}^2+S_{22}^2)},
\end{equation}
where $S_{\alpha\beta}$ is the local shear-rate tensor, expressed as:
\begin{equation}
S_{\alpha \beta}=-\dfrac{1}{2\rho c_s^2\tau^+} \left(\sum_{l=0}^8 (f_l-f_l^{eq})e_{l\alpha}e_{l\beta}+\dfrac{1}{2}(u_\alpha F_\beta+u_\beta F_\alpha)\right),
\label{tenseur_sij}
\end{equation}
where $\tau^+$ is time-dependent in non-Newtonian simulations. According to $\nu=c_s^2(\tau^+ - 1/2)$ and equation (\ref{HB}), $\tau^+$ is updated by:
\begin{equation}
    \begin{array}{c}
         \tau^+ = \dfrac{\sigma_0\dot\gamma^{-1}+K\dot\gamma^{n-1}}{\rho c_s^2}+\dfrac{1}{2}.
    \end{array}{}
    \label{tau_hb}
\end{equation}

To avoid excessive viscosities during simulations and to improve the numerical stability, the Herschel-Bulkley law is truncated. The maximum value of relaxation time $\tau_{\rm max}^+$ is 50 and the viscosity ratio $\mu_{\rm max}/\mu_{\rm min}$ is 1000. A minimum $\dot\gamma$ is set as $10^{-14}$ to avoid zero $\dot\gamma$ in the simulation. As this threshold may seem arbitrary, a larger threshold ($10^{-5}$) was tested. The contours of viscosity and shear rate are very similar to those when the threshold is $10^{-14}$ and the conclusions are unchanged. The viscosity will diverge as the shear rate approaches zero. However, this is reasonable since the mucus is a yield stress fluid. The reader is referred to \citet{GsellDOrtona21} and \citet{GalkoGsell22} for more details on the Herschel-Bulkley model.

The present model relies on five non-dimensional physical parameters, i.e. the ciliary density $\phi$, the interaction length $\lambda$, the Reynolds number $Re={\rho}U_0D/\mu_0$ ($\mu_0$ is the reference viscosity), the Bingham number $Bn$ and the flow index $n$. $\lambda$ is defined as:
\begin{equation}
    \lambda = \dfrac{\sqrt{\mu_0/\kappa}}{D}.
    \label{lambda}
\end{equation}

A high mucus viscosity favors the diffusion of momentum caused by the ciliary beating while a high PCL friction coefficient $\kappa$ prevents it. Thus, $\lambda$ represents the typical range of influence of the ciliated cells \citep{GsellLoiseau20}. The flow of mucus is caused by the momentum transferred from the tip of the cilia. The momentum diffuses over a larger fluid region when $\lambda$ is high. As previously mentioned, mucus flow is almost uniform and parallel to the epithelium. It can be reasonably assumed that no shear exists in the vertical direction for the present model. A reference shear rate can be defined as $\dot\gamma_0=U_0/D$. The reference viscosity is $\mu_0=K\dot\gamma_0^{n-1}$. The general definition of $Re$ becomes \citep{GsellDOrtona21}:
\begin{equation}
    Re = \dfrac{{\rho}U_0^{2-n}D^n}{K},
    \label{gene_Re}
\end{equation}
where $Re=0.1$ is fixed to prevent inertial effects. $K$ can be obtained from equation (\ref{gene_Re}). $Bn$ is defined as:
\begin{equation}
    Bn = \dfrac{\sigma_0}{K}\left(\frac{D}{U_0}\right)^n,
    \label{bingham}
\end{equation}
where the value of $\sigma_0$ is determined by $Bn$ and can be obtained from equation (\ref{bingham}).

\begin{table}
  \begin{center}
\def~{\hphantom{0}}
  \begin{tabular}{lccc}
       Ciliary density ($\phi$) & $0.7-0.8$ \citep{Staudt2014c26} \\
       Dynamic viscosity of the PCL ($\mu_p$) Pa s & $10^{-3}$ \citep{Button2012periciliary} \\
       Thickness of the PCL ($\delta_p$) m & $10^{-5}$ \citep{Button2012periciliary} \\
       Side length of a ciliated element ($D$) m & $2\times10^{-5}$ \citep{LoiseauGsell20}\\
       Viscosity of the healthy mucus ($\mu$) Pa s & $5\times10^{-3}-5\times10^{-2}$ \citep{LoiseauGsell20} \\
       Yield stress of the mucus ($\sigma_0$) Pa & 0.05 \citep{Jory2022mucus} \\
       Flow index of the mucus ($n$) & 0.15 \citep{Jory2022mucus} \\
       Flow consistency of the mucus ($K$) & $0.28$ \citep{Jory2022mucus} \\
       Normal velocity of the mucus ($U_0$) m/s & $1.783 \times 10^{-4}$ \citep{Morgan2004scintigraphic}\\
  \end{tabular}
  \caption{Experimental measurements of physical properties}
  \label{table01}
  \end{center}
\end{table}

The ranges of $\phi$, $\lambda$, $Bn$ and $n$ are physiological and inspired by experimental measurements. Table \ref{table01} shows the experimental measurements of physical properties of the mucus. During the ciliogenesis in experiments \citep{LoiseauGsell20}, ciliary density increases from 0 to a value of approximately $0.7 \sim 0.8$ (normal ciliary density in the airway \citep{Staudt2014c26}). In the present study, $\phi$ varies in the range $0.1\leq \phi \leq 0.7$. From dimensional analysis, the PCL friction coefficient is $\kappa \approx \mu_p/\delta_p^2$, where $\mu_p$ and $\delta_p$ are the dynamic viscosity and thickness of the periciliary layer. $\mu_p \approx 10^{-3}$ Pa s and $\delta_p \approx 10^{-5}$ m are obtained from \citet{Button2012periciliary}. $D \approx 2\times10^{-5}$ m and $\mu \approx 5\times10^{-3} - 5\times10^{-2}$ Pa s (the viscosity of the healthy mucus) are obtained from \citet{LoiseauGsell20}. Therefore, $\lambda$ varies approximately in the range $1\leq \lambda \leq 4$. Yield stress $\sigma_0 \approx 0.05$ Pa, averaged flow index $n \approx 0.15$ and flow consistency $K \approx 0.28$ are obtained and derived from \citet{Jory2022mucus}. The normal mucus velocity $U_0$ is about $1.783 \times 10^{-4}$ m/s \citep{Morgan2004scintigraphic}. Therefore, $Bn$ approximates 0.128, which is very close to the critical $Bn$ (0.15) in the present study for the transition to the fully unidirectional (FU) regime when $\phi=0.7$. In the present study, $Bn$ varies in the range $0\leq Bn \leq 0.3$, enabling the observation of the transition to the FU regime for a very low ciliary density ($\phi=0.1$). In the present study, $n$ varies in the range $0.3\leq n \leq 1$, which is sufficient to observe the transition to the FU regime for a very low ciliary density. Details of the regime transition will be discussed latter. $Bn=0$ and $n=1$ represent a Newtonian case.

About 4500 simulations were performed to produce the results. The general procedure of the present numerical algorithm for the simulation of a coupled cilia-mucus system in Herschel-Bulkley flows can be summarized as follows (the time march loop is performed until the steady solution is obtained): (i) At the $n$-th time step, calculate the angular difference $\Delta\theta^j$ and update the orientation $\theta_c^j$ of $\boldsymbol{F}_c$ by equation (\ref{updatetheta}). (ii) Perform the collision step on the right side of equation (\ref{discret_lbm_force}). Update the value of the relaxation time  $\tau^+$ by equation (\ref{tau_hb}). (iii) Perform the streaming step on the left side of equation (\ref{discret_lbm_force}) to obtain the new $f_l$. (iv) Calculate the new mucus density $\rho$ and mucus velocity $\boldsymbol u$ by equations (\ref{density}) and (\ref{momentum_cor}). Update the value of the frictional force $\boldsymbol{F}_\nu$.

\section{Results and discussion}\label{sec:results_and_discussion}
\subsection{Mucus flow regimes of the cilia-mucus system}

\begin{figure}
\centerline{\includegraphics[width=\linewidth]{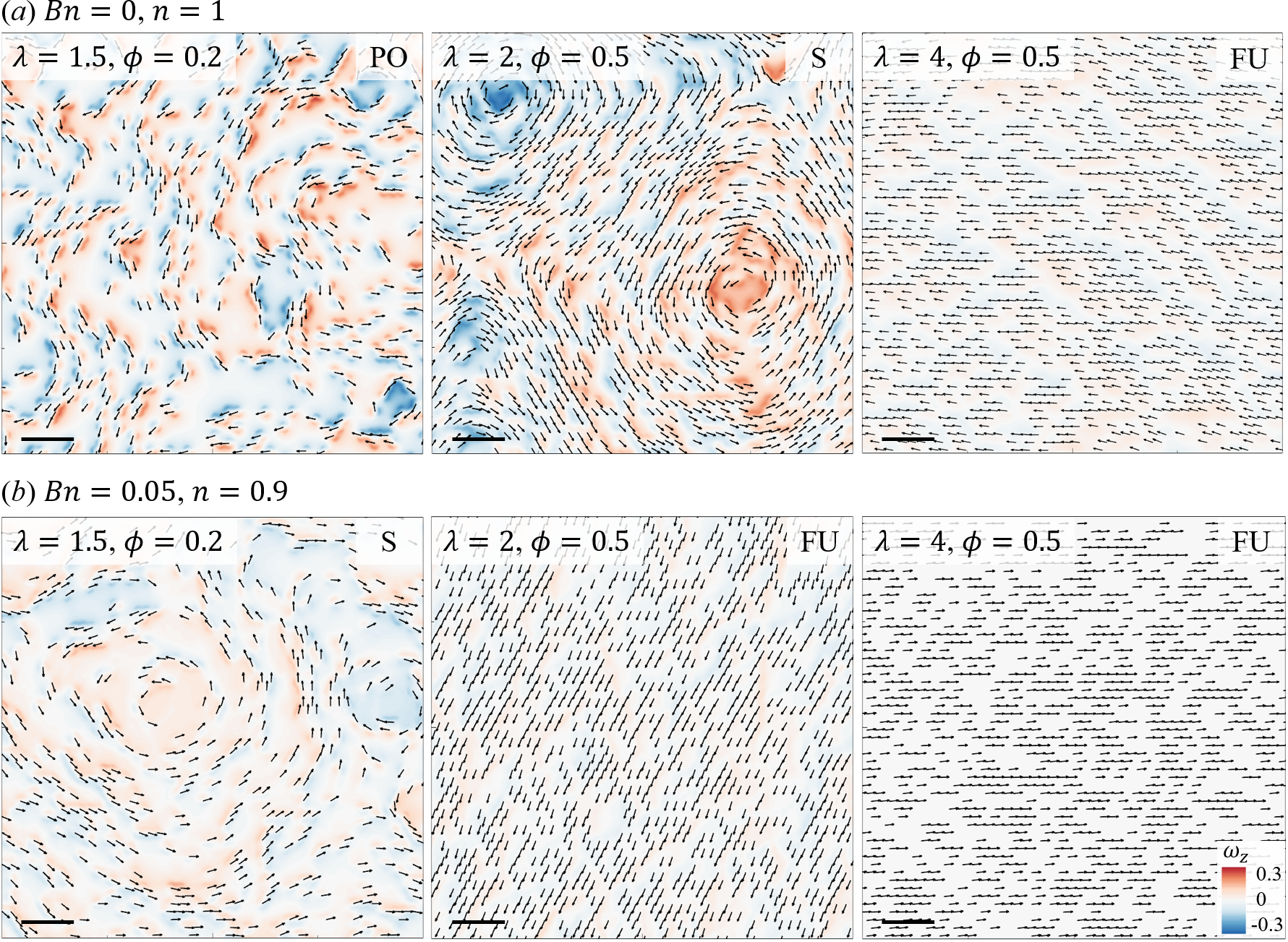}}
\caption{Final steady contours of non-dimensional vorticity $\omega_z$ (vectors indicate ciliary beating direction and the color bar indicates the magnitude of the vorticity) for ($a$) Newtonian fluid (Bingham number $Bn=0$ , flow index $n=1$). Three different mucus flow regimes are shown, from left to right: poorly organized (PO) regime, swirly (S) regime, and fully unidirectional (FU) regime. ($b$) Non-Newtonian fluid ($Bn=0.05$, $n=0.9$). Part of the computational domain is shown and the scale bars correspond to $10D$.}
\label{figure02}
\end{figure}

\noindent In the present study, three distinct mucus flow regimes are observed: a poorly organized (PO) regime, a swirly (S) regime, and a fully unidirectional (FU) regime. Figure \ref{figure02}($a$) shows the contours of non-dimensional vorticity ($\omega_z=(D/U_0)|\nabla\times \boldsymbol u|$) of the three regimes for different $\lambda$ and $\phi$ for a Newtonian fluid ($Bn=0$, $n=1$), where vectors indicate the direction of ciliary beating (local flow). The results in figures \ref{figure02}($a$) and \ref{figure04} obtained by the present model are almost the same as those obtained by \citet{GsellLoiseau20}. The PO regime is characterized by short-range coordination between adjacent cilia without the appearance of large-scale flow structures. The S regime is characterized by long-range coordination of cilia with the formation of obvious mucus swirls. In addition, the FU regime is characterized by long-range coordination of cilia with almost unidirectional flows. These three regimes have been observed in experiments \citep{LoiseauGsell20}. The PO regime corresponds to the pattern in figures 1$(c)-(e)$. The S regime corresponds to the pattern characterized by a small swirl (figures 1$(f)-(h)$). The FU regime corresponds to the pattern characterized by a large swirl that occupies the entire culture chamber (figure 1($i$)), which is caused by the closed culture chamber. The appearance of the unidirectional flow in the present FU regime is mainly attributed to the periodic boundary condition.

\begin{figure}
\centerline{\includegraphics[width=\linewidth]{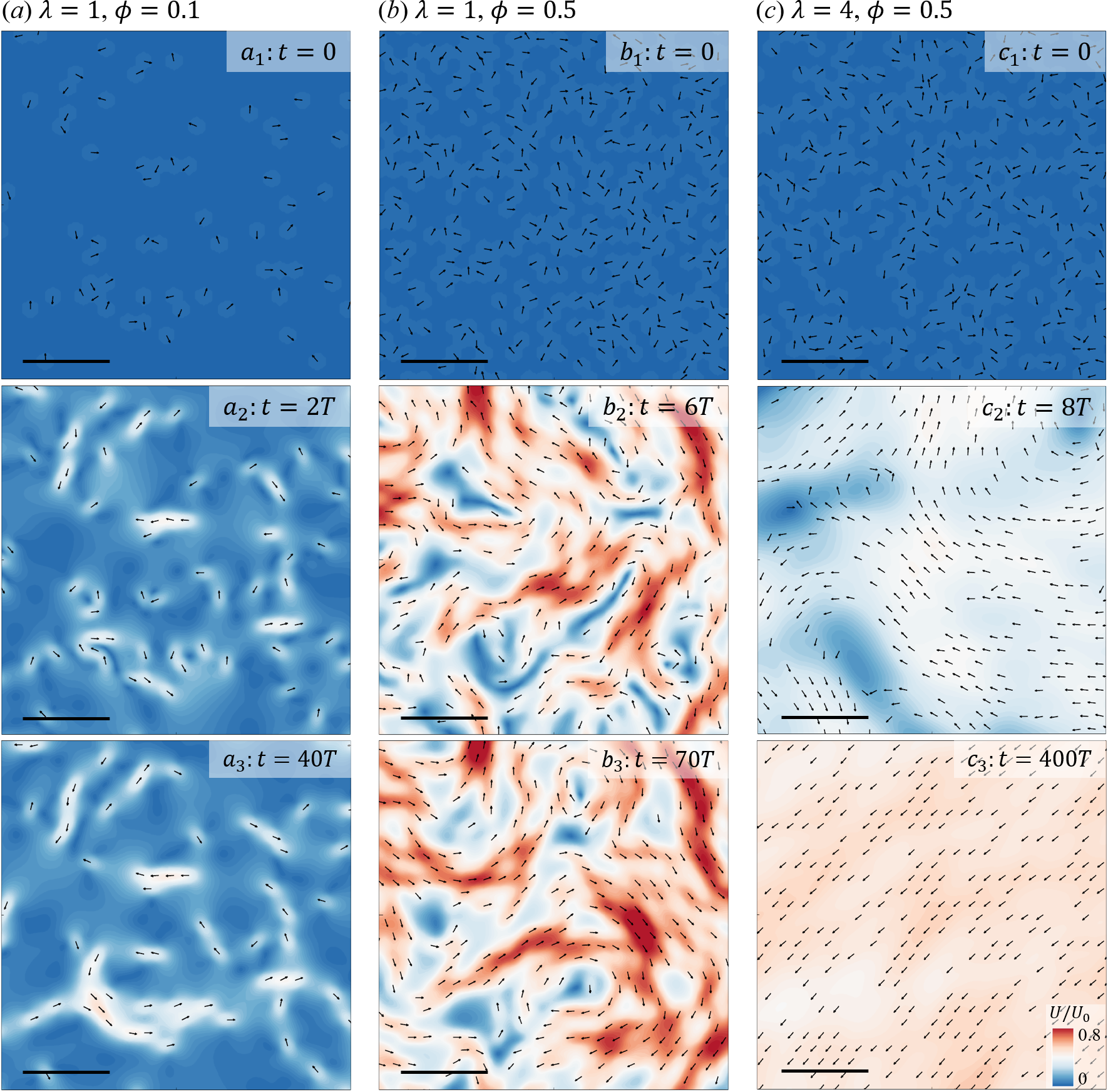}}
\caption{The sequential processes of evolution of the different mucus flow regimes ($Bn=0$, $n=1$, $T=250\Delta t$): ($a$) PO regime for $\lambda=1$ (interaction length) and $\phi=0.1$, ($b$) S regime for $\lambda=1$ and $\phi=0.5$, and ($c$) FU regime for $\lambda=4$ and $\phi=0.5$. Instantaneous contours of the non-dimensional velocity $U/U_0$ are shown. Part of the computational domain is shown and the scale bars correspond to $10D$.}
\label{figure03}
\end{figure}

To examine the formation of the three regimes for different $\lambda$ and $\phi$ ($Bn=0$, $n=1$), the sequential processes of regime evolution from the initial state to the final steady state are plotted in figure \ref{figure03}. The contours are colored by the non-dimensional flow velocity $U/U_0$. Figure \ref{figure03}($a$) shows the formation of the PO regime when $\lambda$ and $\phi$ are very small ($\lambda=1$ and $\phi=0.1$). At instant $a_1$, the domain is initialized with zero flow velocity, random ciliary-beat orientation and random cilia distribution. Mucus flow around the cilia is driven by the ciliary beating, visible at instant $a_2$. The momentum caused by a ciliated element decays rapidly in space due to the small $\lambda$. The ciliated elements are scattered with large distances due to the small $\phi$. Accordingly, the mucus flow caused by different cilia can only interact with each other if they are adjacent, resulting in several local flows without a typical flow structure at instant $a_3$. The flow velocity in the PO regime is very low.

The S regime is formed when $\phi$ is large as shown in figure \ref{figure03}($b$). Mucus flows induced by ciliated elements have the same extension ($\lambda=1$). However, as the ciliary density is higher, the coordination with neighboring ciliated elements is improved and the flow is organized over a greater distance. Several high velocity regions are observed at instant $b_2$ due to the constructive interaction between the adjacent mucus flows. A uniform flow is not formed due to the low $\lambda$. At instant $b_3$, swirls appear after a longer period of coordination than the local flows in the PO regime.

The FU regime is obtained when $\lambda$ is increased as shown in figure \ref{figure03}($c$). The momentum generated by a ciliated element can propagate over a much greater distance. First, at instant $c_2$, swirls are quickly formed due to the rapidly diffused mucus flows. Beyond $c_2$, the further diffused mucus flows influence the ciliary beating over a larger area. After a long period of coordination, the cilia are almost aligned in the same direction, inducing a unidirectional and uniform flow at instant $c_3$.

\subsection{Effects of ciliary density and interaction length}

\begin{figure}
\centerline{\includegraphics[width=\linewidth]{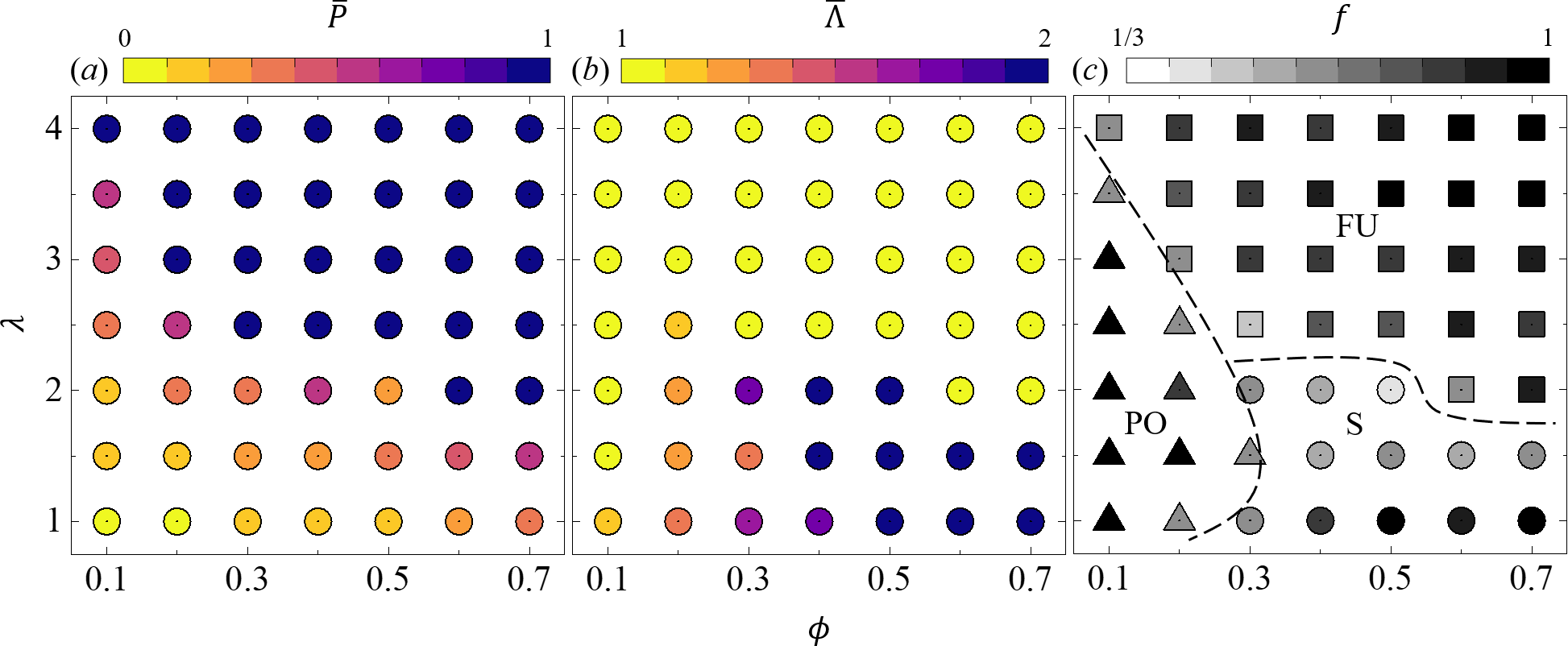}}
\caption{Mucus flow regime diagram depending on $\lambda$ and $\phi$ for $Bn=0$ and $n=1$. Symbols are colored using the values of ($a$) averaged polarization $\bar{P}$, ($b$) averaged integral length $\bar{\Lambda}$, and ($c$) occurrence frequency $f$ of the mucus flow regime; the triangle, circle, and square symbols correspond to PO, S, and FU regimes, respectively.}
\label{figure04}
\end{figure}

\noindent To quantitatively identify the PO, S, and FU regimes in a wide range of parameters, two physical quantities are employed according to the characteristics of the three regimes. The first quantity is the polarization $P$ used to identify the FU regimes, which is the spatial averaging of the unitary velocity vectors, expressed as:
\begin{equation}
    P = \left|\overline{\left(\frac {\boldsymbol{u}}{|\boldsymbol{u}|}\right)}\right|,
\label{P}
\end{equation}
where $P \approx 1$ represents a unidirectional flow. In the present study, $P \geq 0.9$ indicates the FU regime. This critical value is selected based on the observation that $P$ increases sharply from a value below 0.6 to a value above 0.9 when the FU regime appears. This will be discussed in more detail later. The second quantity is the non-dimensional integral length $\Lambda$ normalized by the dimensional interaction length $\sqrt{\mu/\kappa}$:
\begin{equation}
    \Lambda=\sqrt{\frac{\kappa}{\mu}}\int_{0}^{L/2}\frac{R_{x}(\tau)+R_{y}(\tau)}{2}d\tau,
\label{Lambda}
\end{equation}
where $L$ is the length of the computational domain. $R_x(\tau)$ and $R_y(\tau)$ are the $x$ and $y$ components of the auto-correlation functions of the vorticity, respectively:
\begin{equation}
    R_x(\tau)=\frac{\overline{\omega_z(x,y)\omega_z(x+\tau,y)}}{\overline{\omega_z^2}},
\label{Rx}
\end{equation}
\begin{equation}
    R_y(\tau)=\frac{\overline{\omega_z(x,y)\omega_z(x,y+\tau)}}{\overline{\omega_z^2}}.
\label{Ry}
\end{equation}

For $\tau > L/2$, the values of $R_x(\tau)$ and $R_y(\tau)$ are very small, except when $\tau$ approaches $L$ because of the periodic boundary condition. Therefore, the domain of integration is $\tau \in [0,L/2]$ in equation (\ref{Lambda}). $\Lambda$ represents the length scale of flow structures. In particular, the length scale of flow structures is equivalent to the range of influence of the ciliated cells when $\Lambda=1$. Small and large $\Lambda$ indicate the PO and the S regimes, respectively. The increase in $\Lambda$ is relatively smooth as the PO regime transitions to the S regime. The critical value of $\Lambda=1.5$ is selected based on the observation of the flow regime from numerous simulations. In summary, $P < 0.9$ and $\Lambda < 1.5$ indicate the PO regime, $P < 0.9$ and $\Lambda \geq 1.5$ indicate the S regime, $P \geq 0.9$ indicates the FU regime.

For comparison, we first examine the effects of $\lambda$ and $\phi$ on the formation of the mucus flow regime in the Newtonian case ($Bn=0$, $n=1$). Figure \ref{figure04} shows a phase diagram in the ranges $1 \leq \lambda \leq 4$ and $0.1 \leq \phi \leq 0.7$. Random initialization can result in different flow regimes under certain conditions. Therefore, for each case in the diagram, 20 randomly initialized simulations were performed. $f$ is the occurrence frequency of the most frequent flow regime over a set of 20 simulations. $\bar{P}$ and $\bar{\Lambda}$ are the averaged polarization and integral length calculated by the simulations converged to the most frequent flow regime. In figure \ref{figure04}($a$), symbols are colored by the value of $\bar{P}$. $\bar{P}$ increases with increasing $\lambda$ and $\phi$. In figure \ref{figure04}($b$), symbols are colored by the value of $\bar{\Lambda}$. For $\bar{P} \geq 0.9$ (FU regime), $\bar{\Lambda}$ is set to be 1 and has no physical meaning. We do not discuss the length scale of the flow structures for the FU regime because it is theoretically infinite. $\bar\Lambda$ increases with $\phi$. The effect of $\lambda$ is less clear, but a small tendency of increasing $\bar\Lambda$ with decreasing $\lambda$ may be observed. In figure \ref{figure04}($c$), symbols are colored by the value of $f$. The diagram is divided into three regions by dashed lines according to the maps of $\bar{P}$ and $\bar{\Lambda}$. The PO regime appears in the region with low $\lambda$ and $\phi$. The S regime appears in the region with low $\lambda$ and high $\phi$. The FU regime appears in the region with high $\lambda$ and high $\phi$. As mentioned in figure \ref{figure03}, these are mainly determined by the range of influence of the ciliated cells and the interaction of the mucus flows caused by adjacent cilia. $f$ is relatively small for the points near the regime boundary.

\subsection{Effects of yield stress and shear thinning properties}

\begin{figure}
\centerline{\includegraphics[width=\linewidth]{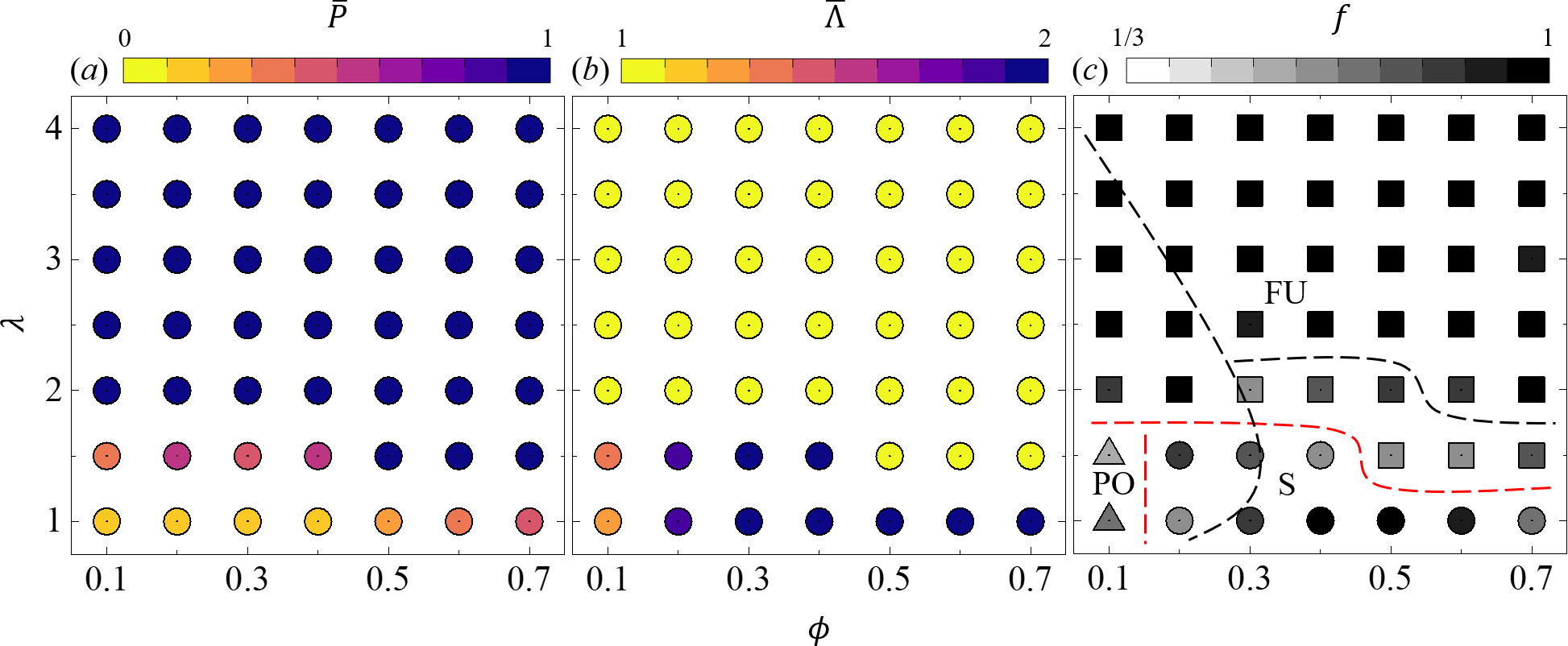}}
\caption{Mucus flow regime diagram depending on $\lambda$ and $\phi$ for $Bn=0.05$ and $n=0.9$. Symbols are colored using the values of ($a$) $\bar{P}$, ($b$) $\bar{\Lambda}$, and ($c$) $f$. The red and black dashed lines represent the boundaries of the non-Newtonian cases ($Bn=0.05$, $n=0.9$) and the Newtonian cases ($Bn=0$, $n=1$), respectively.}
\label{figure05}
\end{figure}

\noindent For studying the effects of non-Newtonian properties on the flow regime formation, a phase diagram in the ranges $1 \leq \lambda \leq 4$ and $0.1 \leq \phi \leq 0.7$ is shown in figure \ref{figure05}. The simulated mucus is shear thinning $n=0.9$ and has a yield stress $Bn=0.05$. The regions with high $\bar{P}$ and low $\bar{\Lambda}$ are significantly enlarged compared to those for $Bn=0$ and $n=1$, as shown in figures \ref{figure05}($a$) and ($b$). Figure \ref{figure05}($c$) clearly shows the displacement of the regime boundary, where the red and black dashed lines represent the boundaries of the non-Newtonian cases ($Bn=0.05$, $n=0.9$) and the Newtonian cases ($Bn=0$, $n=1$), respectively. In general, the regions of the PO and S regimes are reduced and the region of the FU regime is increased. The PO regime only appears when $\lambda$ and $\phi$ are very low. The S regime is obtained at lower $\lambda$, while it appears in a wider range of $\phi$. A lower $\lambda$ allows the activation of the FU regime. The boundary between the PO and FU regimes is significantly changed. These indicate that the flow regimes of all cases can be converted to the FU regime by varying $Bn$ and $n$. The point with $\phi=0.1$ and $\lambda=1$ would be the last case to complete this conversion. The flow regime of three non-Newtonian cases is shown in figure \ref{figure02}($b$), which shows a significant regime transition when compared with the Newtonian cases in figure \ref{figure02}($a$).

\begin{figure}
\centerline{\includegraphics[width=\linewidth]{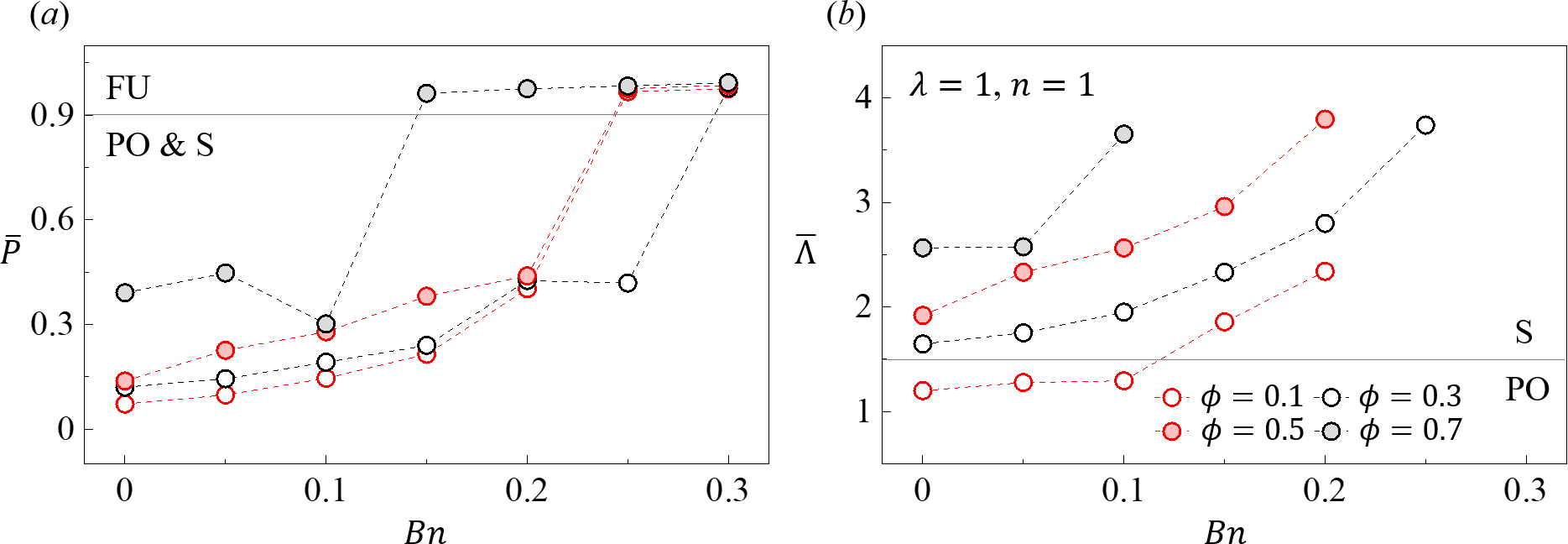}}
\caption{Values of ($a$) $\bar{P}$ and ($b$) $\bar{\Lambda}$ as functions of $Bn$ for different $\phi$ ($\lambda=1$, $n=1$).}
\label{figure06}
\end{figure}

Here we further examine the effects of $Bn$ and $n$ on the flow regime formation in detail. First, $n=1$ is fixed to explore the effect of $Bn$ independently, i.e. a Herschel-Bulkley fluid reduces to a Bingham fluid. In figure \ref{figure06}($a$), the value of $\bar P$ is presented with respect to the Bingham number $Bn$. Recall that to obtain the value of $\bar P$, only the simulations in the most frequent regime have been used. Thus, each curve represents two sets of data that can be considered independently, and the sharp transition indicates the critical value of $Bn$ that induces a transition to the FU regime. In the PO $\&$ S regime, $\bar P$ increases monotonously with $Bn$, except for the case $\phi=0.7$ where the variation is less clear. After the transition to the FU regime, $\bar P$ remains almost constant. In general, a larger $\phi$ leads to a larger $\bar{P}$, confirming the results in figures \ref{figure04}($a$) and \ref{figure05}($a$). The critical $Bn$ increases and then decreases with increasing $\phi$, resulting in a maximum critical $Bn$ at $\phi=0.3$. In figure \ref{figure06}($b$), the variation of $\bar{\Lambda}$ as a function of $Bn$ is shown. The FU regime is not included due to its theoretically infinite length scale of the flow structures. $\bar{\Lambda}$ increases with increasing $Bn$ for different $\phi$. A larger $\phi$ leads to a larger $\bar{\Lambda}$, confirming the results in figures \ref{figure04}($b$) and \ref{figure05}($b$). For $\phi=0.1$, a transition from the PO regime to the S regime is observed by increasing $Bn$. For $\phi > 0.1$, only the S regime is obtained. The above results suggest that the range of influence of the ciliated cells is increased by increasing $Bn$.

\begin{figure}
\centerline{\includegraphics[width=\linewidth]{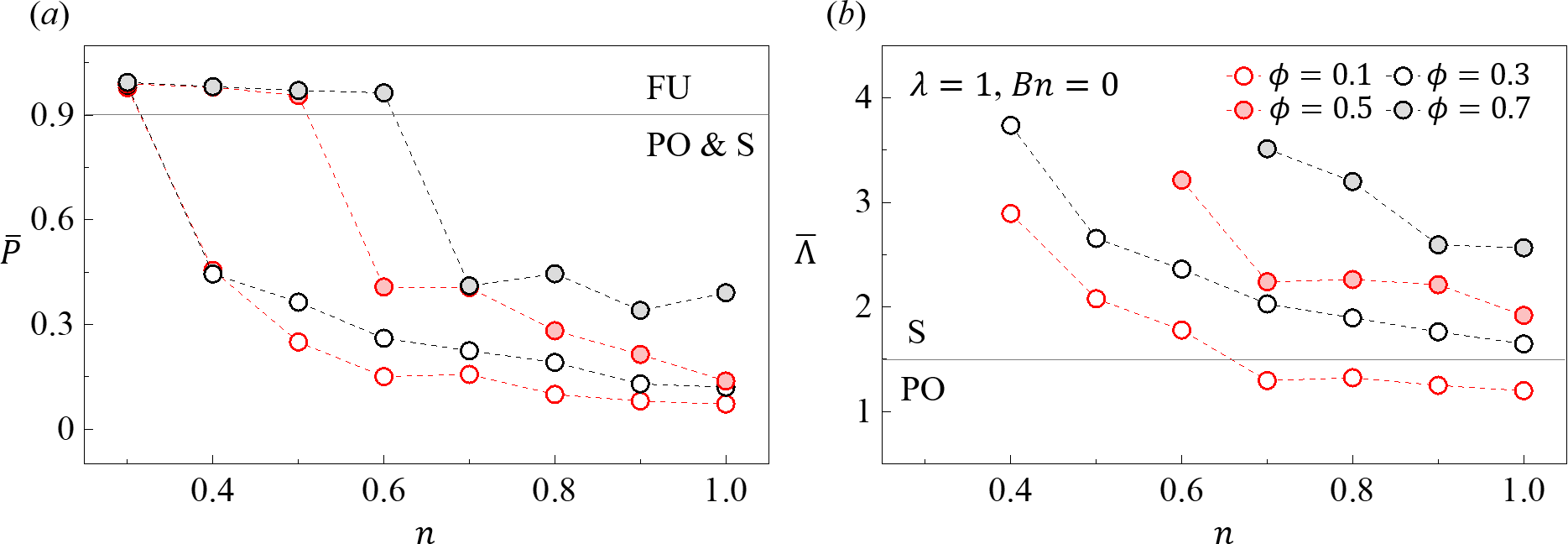}}
\caption{Values of ($a$) $\bar{P}$ and ($b$) $\bar{\Lambda}$ as functions of $n$ for different $\phi$ ($\lambda=1$, $Bn=0$).}
\label{figure07}
\end{figure}

Figure \ref{figure07} shows the variations of $\bar{P}$ and $\bar{\Lambda}$ as functions of $n$ for different $\phi$. $\lambda=1$ and $Bn=0$ are fixed. The sharp increase of $\bar P$ in figure \ref{figure07}($a$) indicates the critical value of $n$ that leads to a transition to the FU regime. The critical $n$ increases with increasing $\phi$. In contrast to the effect of $Bn$, $\bar P$ increases monotonously with decreasing $n$ in the PO $\&$ S regime. After the transition to the FU regime, $\bar P$ remains almost constant. In figure \ref{figure07}($b$), the decrease of $n$ can also lead to a transition from the PO regime to the S regime when the ciliary density is low ($\phi=0.1$). The above results suggest that the range of influence of the ciliated cells is increased by decreasing $n$.

\begin{figure}
\centerline{\includegraphics[width=\linewidth]{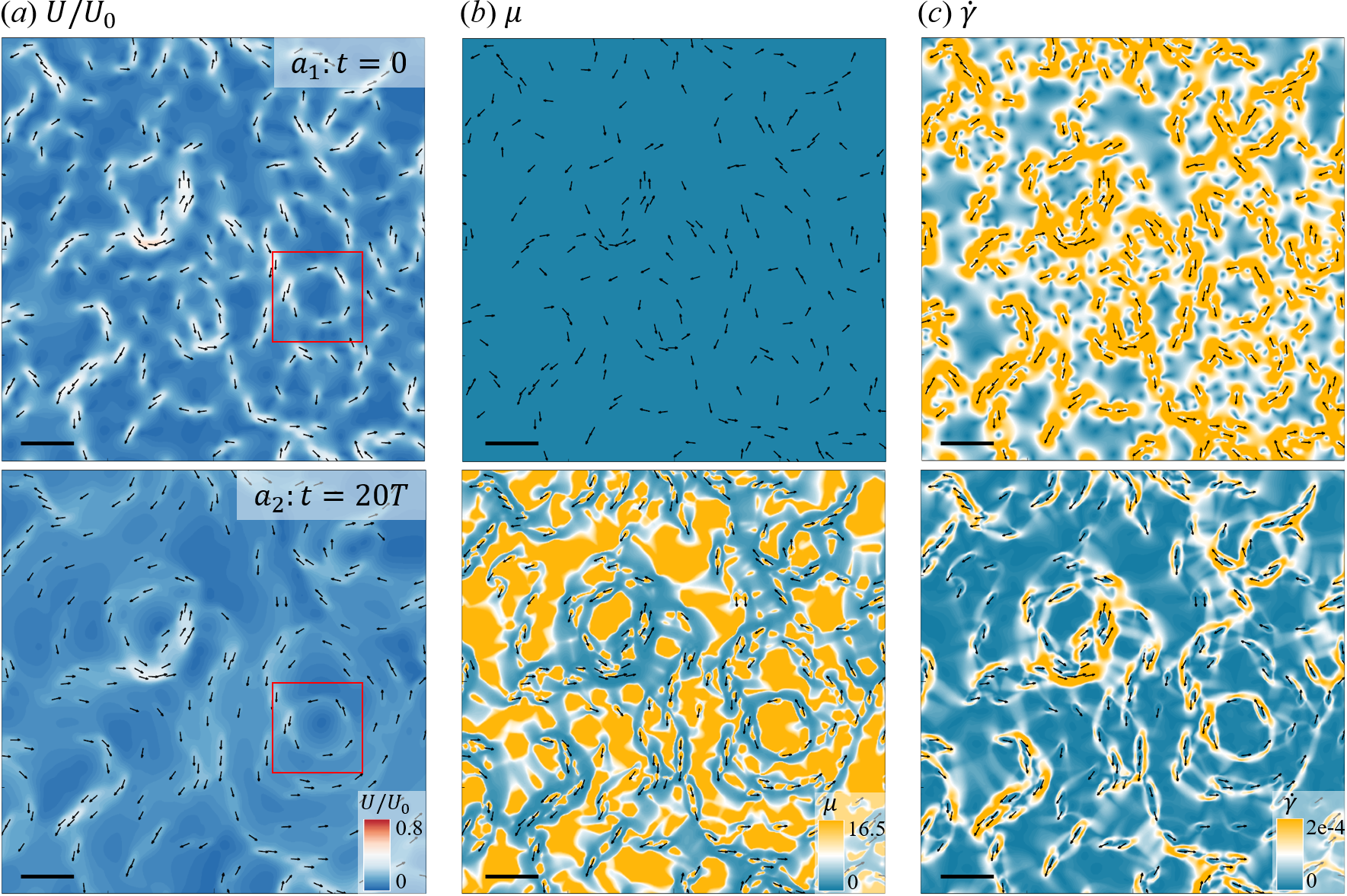}}
\caption{The sequential process of a transition from PO regime (instant $a_1$, $Bn=0$) to S regime (instant $a_2$, $Bn=0.15$) by increasing $Bn$. The steady solution for $Bn=0$ (instant $a_1$) is used as the initial condition for the simulation with $Bn=0.15$. A steady solution for $Bn=0.15$ is obtained at instant $a_2$. Instantaneous contours are colored by ($a$) $U/U_0$, ($b$) dynamic fluid viscosity $\mu$ (its value at instant $a_1$ corresponds to a Newtonian fluid), and ($c$) local shear-rate magnitude $\dot \gamma$ ($\lambda=1$, $\phi=0.1$, $n=1$). Part of the computational domain is shown and the scale bars correspond to $10D$.}
\label{figure08}
\end{figure}

\begin{figure}
\centerline{\includegraphics[width=\linewidth]{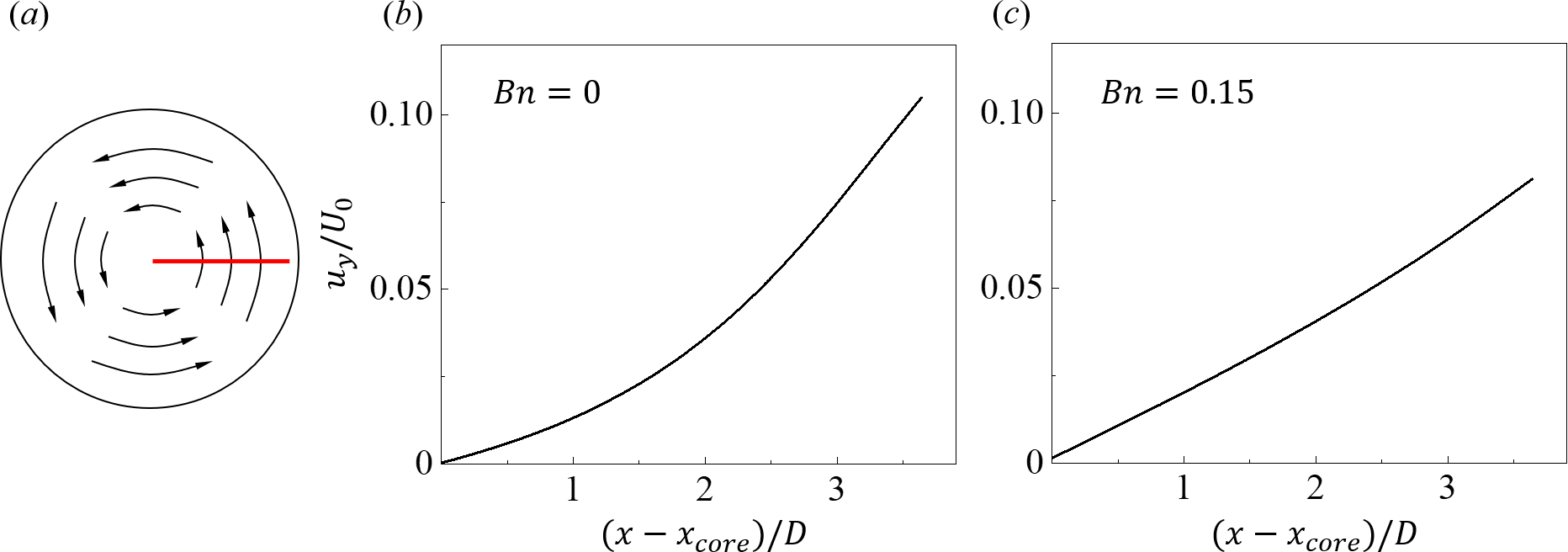}}
\caption{($a$) Schematic of velocity extraction in the core of a swirl. Distribution of the longitudinal velocity $u_y$ along the radial direction in the core of swirls marked in figure \ref{figure08}($a$): ($b$) $Bn=0$, ($c$) $Bn=0.15$.}
\label{figure09}
\end{figure}

To visualize the transition from the PO regime to the S regime by increasing $Bn$, we further examine the instantaneous contours of the flow velocity $U/U_0$, the dynamic fluid viscosity $\mu$, and the local shear rate magnitude $\dot \gamma$ by increasing $Bn$ from 0 to 0.15 in figure \ref{figure08} ($\lambda=1$, $\phi=0.1$, $n=1$). Note that the steady solution for $Bn=0$ (instant $a_1$) is used as the initial condition for the simulation with $Bn=0.15$. A steady solution for $Bn=0.15$ is obtained at instant $a_2$. In figure \ref{figure08}($a$), swirls are more pronounced at instant $a_2$, corresponding to the increase in $\Lambda$ from 6.36 to 10.99. This is mainly caused by the evolution of the $\mu$ and $\dot \gamma$ distributions in figures \ref{figure08}($b$) and ($c$). At instant $a_1$, $\mu$ is close to its reference value and $\dot \gamma$ caused by the ciliary beating is high. At instant $a_2$, $\mu$ in the region with low shear rate significantly increases with increasing $Bn$ (yield stress). This can be verified by checking equation (\ref{HB}). The momentum diffuses farther when the viscosity is high. Accordingly, the mucus flow caused by a ciliated element affects the ciliary-beat orientation further away, which favors the coordination of the cilia, thereby resulting in a transition from the PO regime to the S regime. A reorientation of the cilia can be clearly observed in figure \ref{figure08}(a). However, the high viscosity makes the mucus difficult to shear, resulting in a decrease in $\dot \gamma$ in figure \ref{figure08}($c$). At instant $a_2$, the low $\dot \gamma$ region (blue color) corresponding to the high $\mu$ region and the non-ciliated region are in solid body rotation or in solid body motion. This solid body rotation has been observed experimentally in the core of swirl \citep{LoiseauGsell20}, visualized in figures 1($h$) and ($i$) and supplementary movies 5 and 6 (\url{https://www.biorxiv.org/content/10.1101/2019.12.16.878108v1.supplementary-material}). The role of the yield stress in generating solid body rotation is that the effective viscosity diverges as the shear rate approaches zero. Here we examine the distribution of longitudinal velocity $u_y$ in the core of swirls marked in figure \ref{figure08}($a$) for $Bn=0$ and 0.15. $u_y$ is extracted along the red line as schematized in figure \ref{figure09}($a$). The distributions of $u_y$ are shown in figures \ref{figure09}($b$) and ($c$). $u_y$ varies linearly in the radial direction when $Bn=0.15$, indicating a solid body rotation. This was not reproduced in the previous study \citep{GsellLoiseau20} since a Newtonian fluid was modeled. The addition of the yield stress property to the mucus allows a more accurate modeling of the experiments.

\begin{figure}
\centerline{\includegraphics[width=\linewidth]{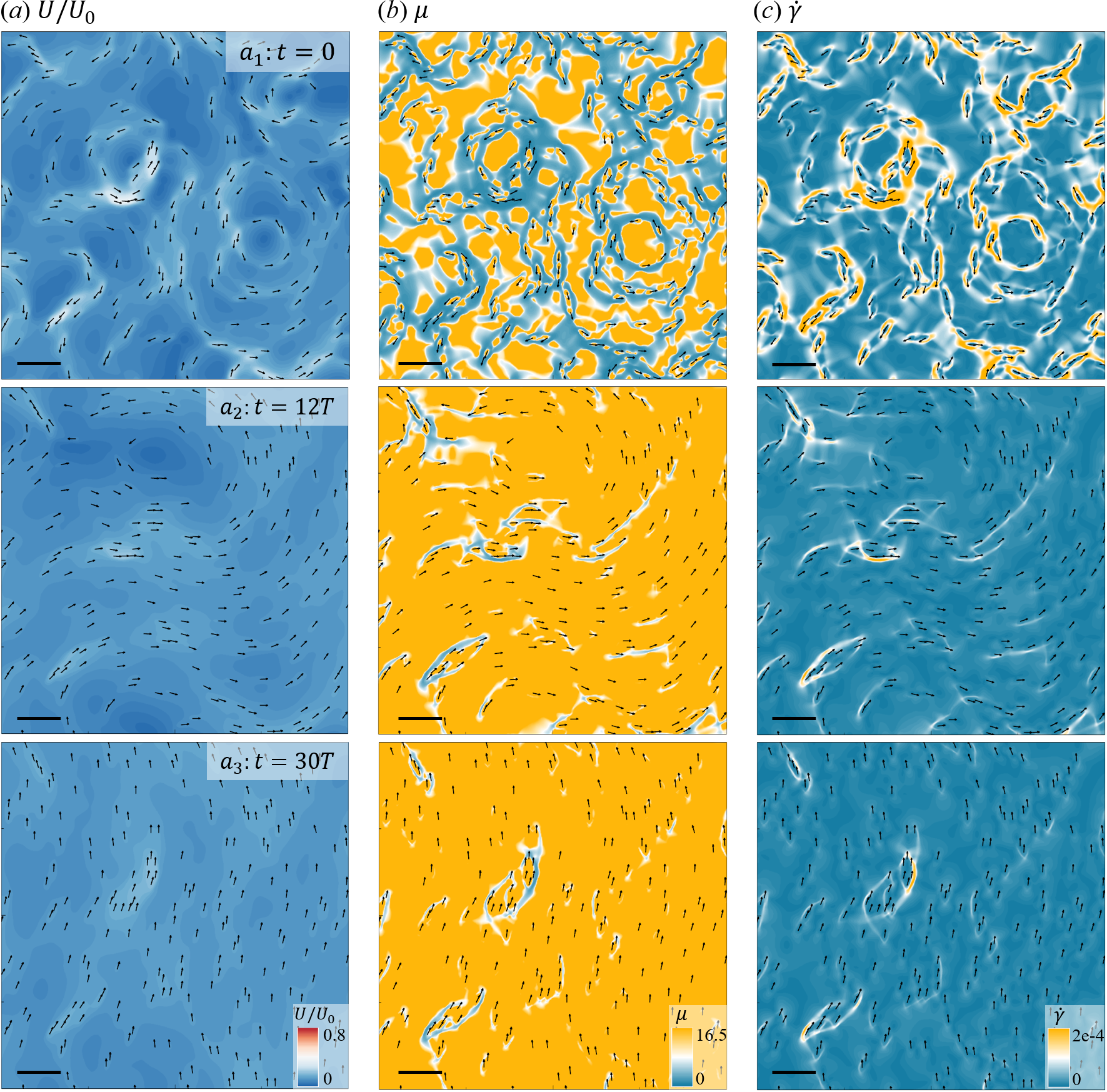}}
\caption{The sequential process of a transition from S regime (instant $a_1$, $Bn=0.15$) to FU regime (instant $a_3$, $Bn=0.3$) by increasing $Bn$: instantaneous contours of ($a$) $U/U_0$, ($b$) $\mu$, and ($c$) $\dot \gamma$ ($\lambda=1$, $\phi=0.1$, $n=1$). Part of the computational domain is shown and the scale bars correspond to $10D$.}
\label{figure_tranBn2}
\end{figure}

The instantaneous contours of $U/U_0$, $\mu$, and $\dot \gamma$ by increasing $Bn$ from 0.15 to 0.3 are shown in figure \ref{figure_tranBn2} ($\lambda=1$, $\phi=0.1$, $n=1$). The steady solution for $Bn=0.15$ (instant $a_2$ in figure \ref{figure08} or instant $a_1$ in figure \ref{figure_tranBn2}) is used as the initial condition for the simulation with $Bn=0.3$. A steady solution for $Bn=0.3$ is obtained at instant $a_3$. The further increase in $Bn$ substantially increases $\mu$ at instant $a_2$ irrespective of the high $\dot \gamma$ region at instant $a_1$. This further enhances the diffusion of momentum and the coordination of different cilia, resulting in a larger swirl in figure \ref{figure_tranBn2}($a$) (instant $a_2$). The increase in $\mu$ decreases $\dot \gamma$, which in turn increases $\mu$ at instant $a_3$. Thus, there is a positive feedback between the increased $\mu$ and the decreased $\dot \gamma$ for the flow with yield stress. The S regime gradually converts to the FU regime due to the further diffusion of momentum.

\begin{figure}
\centerline{\includegraphics[width=\linewidth]{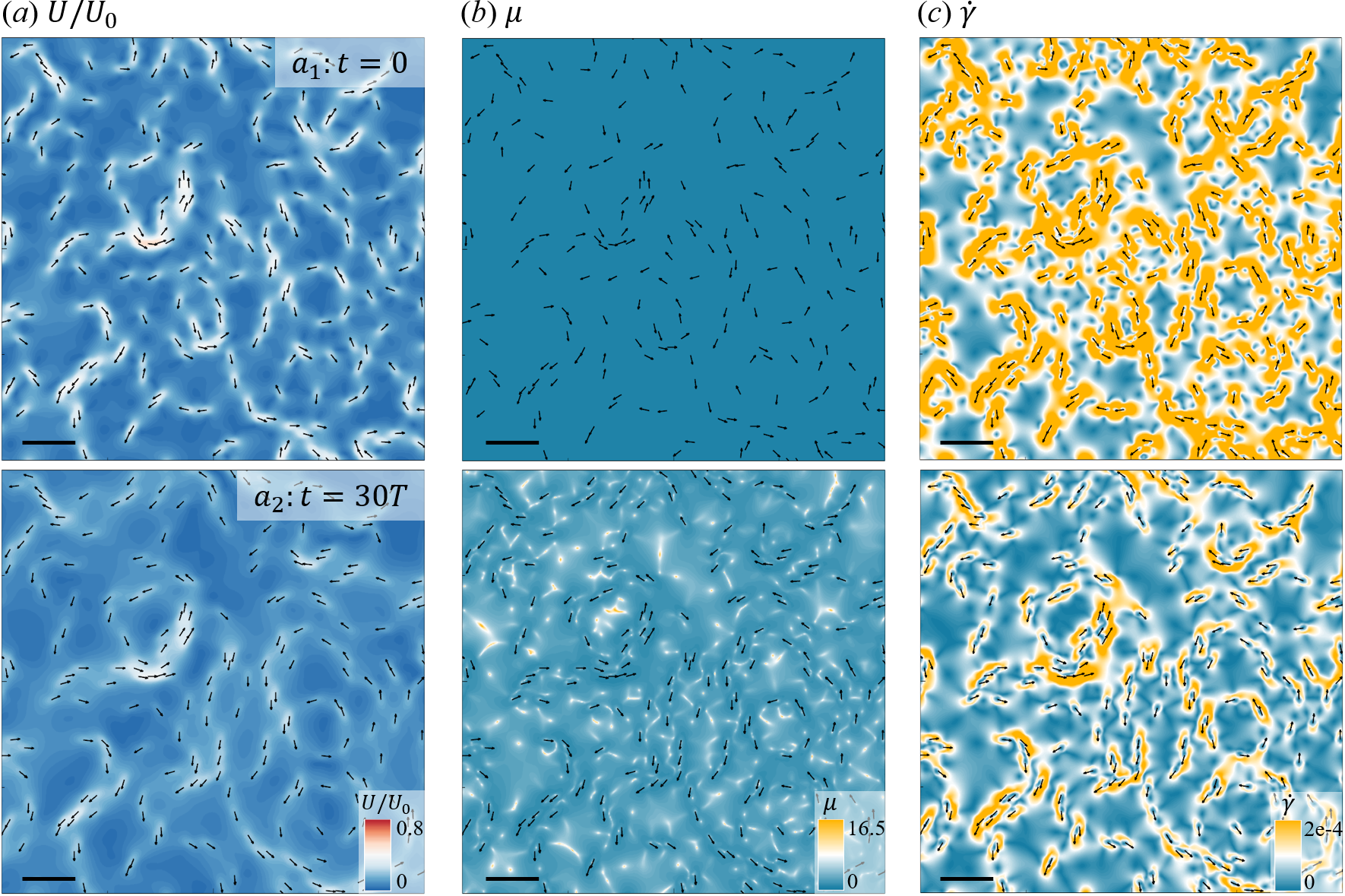}}
\caption{The sequential process of a transition from PO regime (instant $a_1$, $n=1$) to S regime (instant $a_2$, $n=0.6$) by decreasing $n$: instantaneous contours of ($a$) $U/U_0$, ($b$) $\mu$ (its value at instant $a_1$ corresponds to a Newtonian fluid), and ($c$) $\dot \gamma$ ($\lambda=1$, $\phi=0.1$, $Bn=0$). Part of the computational domain is shown and the scale bars correspond to $10D$.}
\label{figure_tranN1}
\end{figure}

For the regime transition induced by varying $n$, the instantaneous contours of $U/U_0$, $\mu$, and $\dot \gamma$ by decreasing $n$ from 1 to 0.6 are shown in figure \ref{figure_tranN1} ($\lambda=1$, $\phi=0.1$, $Bn=0$). According to equations (\ref{HB}) and (\ref{gene_Re}), the viscosity is $\mu=(\rho U_0^2 \dot \gamma^{-1} Re^{-1})(D\dot \gamma/U_0)^n$. $\mu$ is increased when $n$ decreases from 1 to 0.6 ($D\dot \gamma/U_0$ \textless 1), enhancing the diffusion of momentum and reducing $\dot \gamma$. The decrease in $\dot \gamma$ in turn leads to an increase in $\mu$ due to the shear thinning behavior. There is also a positive feedback between the increased $\mu$ and the decreased $\dot \gamma$ for the shear thinning flow. A transition from the PO regime to the S regime is observed at instant $a_2$ due to the enhanced diffusion of
momentum, corresponding to the increase in $\Lambda$ from 6.36 to 10.15. The effect of shear thinning ($n=0.6$ and $Bn=0$) on $\mu$ is weak compared to the effect of yield stress ($n=1$ and $Bn=0.15$).

\begin{figure}
\centerline{\includegraphics[width=\linewidth]{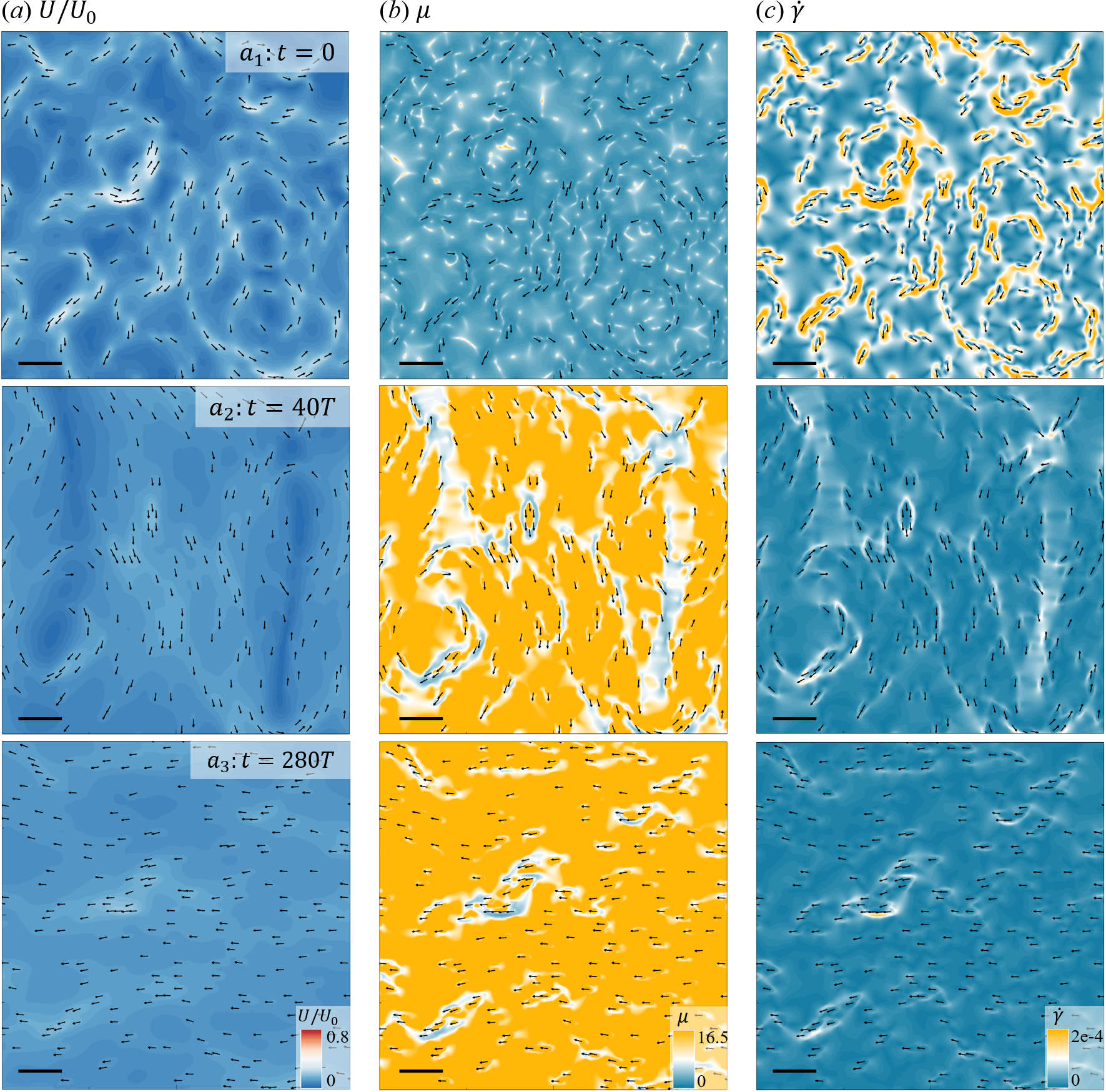}}
\caption{The sequential process of a transition from S regime (instant $a_1$, $n=0.6$) to FU regime by decreasing $n$ (instant $a_3$, $n=0.3$): instantaneous contours of ($a$) $U/U_0$, ($b$) $\mu$, and ($c$) $\dot \gamma$ ($\lambda=1$, $\phi=0.1$, $Bn=0$). Part of the computational domain is shown and the scale bars correspond to $10D$.}
\label{figure_tranN2}
\end{figure}

Figure \ref{figure_tranN2} shows the instantaneous contours of $U/U_0$, $\mu$, and $\dot \gamma$ by further decreasing $n$ from 0.6 to 0.3 ($\lambda=1$, $\phi=0.1$, $Bn=0$). As mentioned above, the decrease of $n$ leads to the increase of $\mu$. The substantial enhancement of shear thinning effect significantly enhance the positive feedback between the increased $\mu$ and the decreased $\dot \gamma$. At instant $a_2$, the increase of $\mu$ results in the formation of large scale swirls. At instant $a_3$, the further increase of $\mu$ and the full coordination of cilia and mucus induce a transition from the S regime to the FU regime. In summary, both the increase of $Bn$ and the decrease of $n$ lead to the successive appearance of PO, S, and FU regimes. This is closely related to the increase of $\mu$ and the diffusion of momentum. Note that the flow velocity in the ciliated region are lower for the FU regime than for the PO (or S) regime. In the FU regime, the momentum diffuses into the non-ciliated region, and the flow velocity is averaged over ciliated nodes with beating force and non-ciliated nodes with only friction. In the PO regime, the velocity remains localized above the ciliated region, which is ineffective for the mucus transport.

\begin{figure}
\centerline{\includegraphics[width=\linewidth]{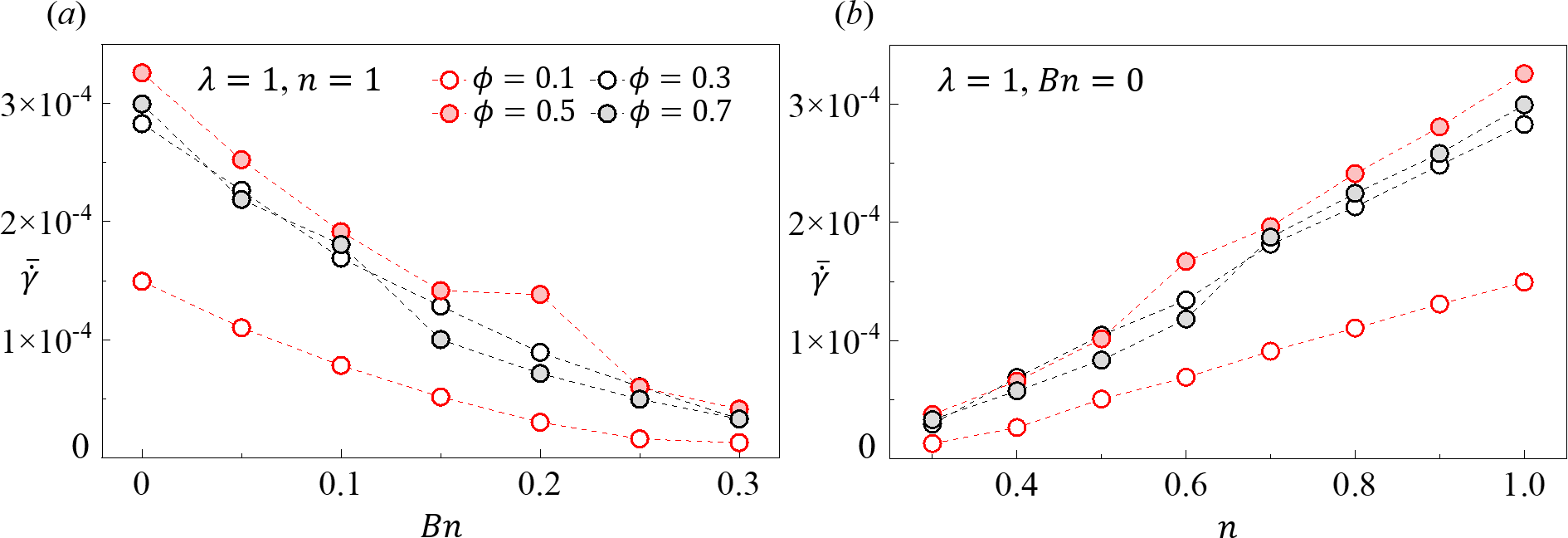}}
\caption{Averaged shear-rate magnitude $\bar {\dot \gamma}$ as a function of ($a$) $Bn$ ($\lambda=1$, $n=1$) and ($b$) $n$ ($\lambda=1$, $Bn=0$) for different $\phi$.}
\label{figure_curveShear}
\end{figure}

The variation of $\dot \gamma$ directly influences $\mu$ under the yield stress and shear thinning effects. Here we calculate the spatially averaged shear-rate magnitude $\bar {\dot \gamma}$ for different cases. The variation of $\bar {\dot \gamma}$ as a function of $Bn$ ($\lambda=1$, $n=1$) and $n$ ($\lambda=1$, $Bn=0$) for different $\phi$ is shown in figure \ref{figure_curveShear}. $\bar {\dot \gamma}$ decreases with increasing $Bn$ and decreasing $n$, indicating the increase in mucus viscosity. In addition, $\bar {\dot \gamma}$ increases with increasing $\phi$ until $\phi=0.5$ due to the increase in the number of cilia. The further increase in $\phi$ significantly enhances the coordination between different cilia, which tend to beat in the same direction and result in a lower shear. Beyond $\phi=0.3$, $\bar {\dot \gamma}$ is relatively insensitive to the increase in $\phi$.

\subsection{Effective interaction length}

\begin{figure}
\centerline{\includegraphics[width=\linewidth]{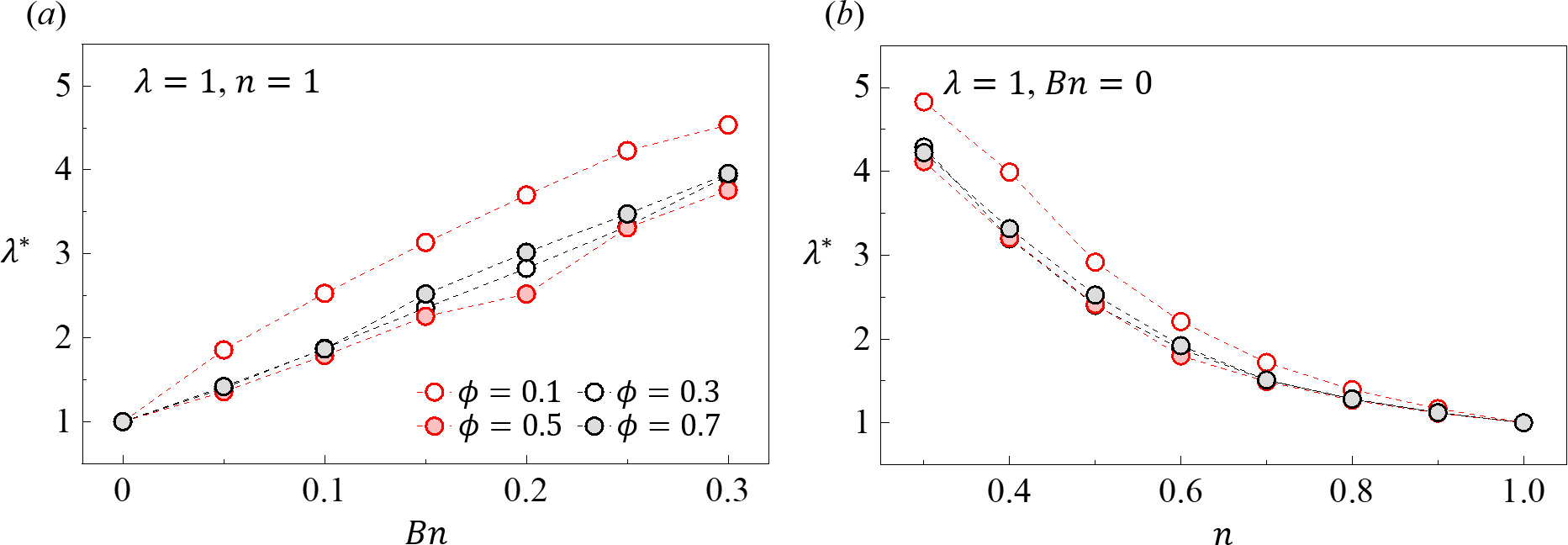}}
\caption{Effective interaction length $\lambda^*$ as a function of ($a$) $Bn$ ($\lambda=1$, $n=1$) and ($b$) $n$ ($\lambda=1$, $Bn=0$) for different $\phi$.}
\label{figure_curveEfLam}
\end{figure}

The mucus viscosity $\mu$ is the dominant parameter that affects the regime formation when varying $Bn$ and $n$ according to the above discussions. In the present study, the viscosity effect is included in $\lambda$, which is defined based on the reference mucus viscosity. To consider the variation of $\mu$, a spatially averaged viscosity $\bar{\mu}$ is calculated for the cases with different $Bn$ and $n$. The lattice nodes with maximum relaxation time are excluded due to the truncated Herschel-Bulkley law used in the present study. In fact, their viscosity should be considered almost infinite. An effective interaction length $\lambda^*$ is defined based on $\bar{\mu}$ instead of $\mu$. $\lambda^*$ is found to be more suitable to represent the range of influence of the ciliated cells. Figure \ref{figure_curveEfLam} shows the variation of $\lambda^*$ as a function of $Bn$ ($\lambda=1$, $n=1$) and $n$ ($\lambda=1$, $Bn=0$) for different $\phi$. $\lambda^*$ increases with increasing $Bn$ and decreasing $n$, which favors the diffusion of momentum and the coordination between cilia and mucus, thereby resulting in the regime transition. This confirms the results shown in figures \ref{figure08} and \ref{figure_tranBn2}-\ref{figure_tranN2}. For $\phi \geq 0.3$, $\lambda^*$ is smaller than that for $\phi=0.1$, and the curves of $\lambda^*$ almost collapse onto a single curve. This corresponds to the variation of $\bar {\dot \gamma}$ in figure \ref{figure_curveShear}. Furthermore, the variation of $\lambda^*$ is opposite to the variation of $\bar {\dot \gamma}$ due to the yield stress and shear thinning effects.

\begin{figure}
\centerline{\includegraphics[width=\linewidth]{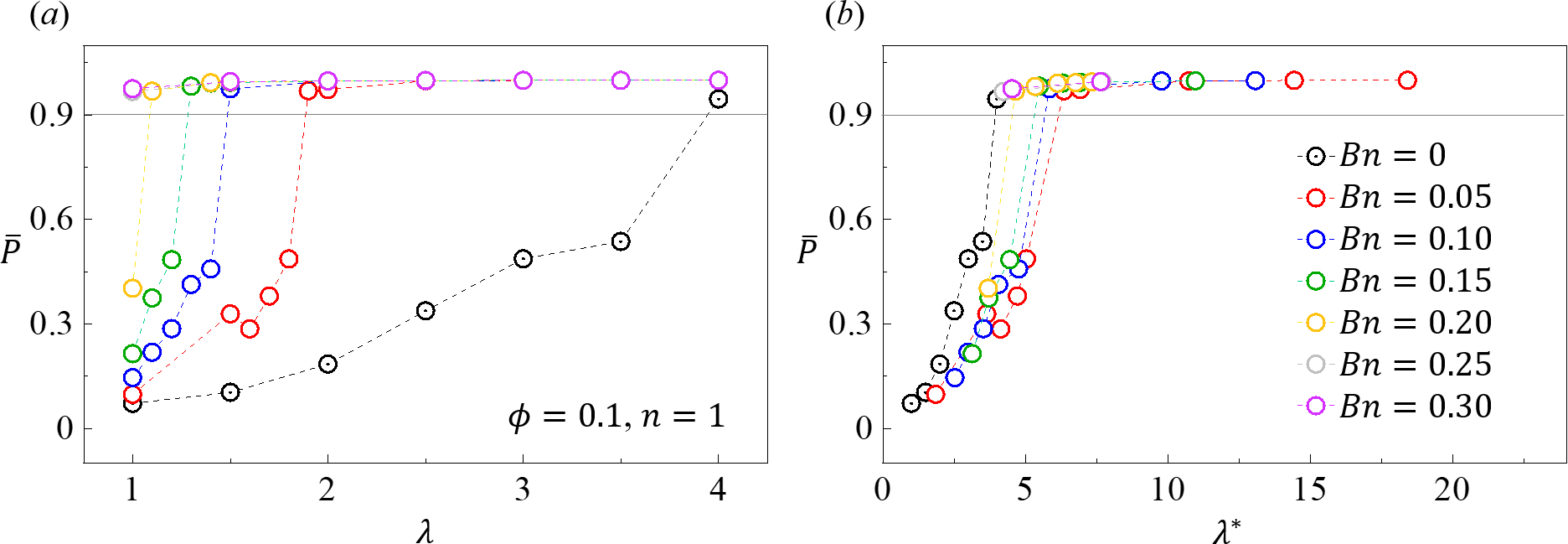}}
\caption{The value of $\bar P$ as a function of ($a$) $\lambda$ and ($b$) $\lambda^*$ for different $Bn$ ($\phi=0.1$, $n=1$).}
\label{figure_curvePvsLamN}
\end{figure}

To examine the dependence of the regime formation on $\lambda^*$, we compare the variation of $\bar P$ as a function of $\lambda$ and $\lambda^*$ for different $Bn$ ($\phi=0.1$, $n=1$) in figure \ref{figure_curvePvsLamN}. Recall that to obtain the value of $\bar P$, only the simulations in the most frequent regime have been used. $\bar P$ increases with increasing $\lambda$ and a rapid increase of $\bar P$ can be observed at the critical points for the appearance of the FU regime. The FU regime appears at a smaller $\lambda$ as $Bn$ is increased. Here we only consider the critical condition of FU regime formation because the FU regime is the most efficient for mucus transport. The curves of $\bar P$ are scattered for different $Bn$ in figure \ref{figure_curvePvsLamN}($a$). In figure \ref{figure_curvePvsLamN}($b$), the curves collapse onto a single curve as a whole for $Bn > 0$. Here, $Bn=0$ deviates significantly from the collapsed curve. $\lambda^*$ is not enough to predict the critical condition for different $Bn$. This may be related to the fact that the increase in $Bn$ leads to a sharp increase in $\mu$, which is truncated when the maximum value is reached.

\begin{figure}
\centerline{\includegraphics[width=\linewidth]{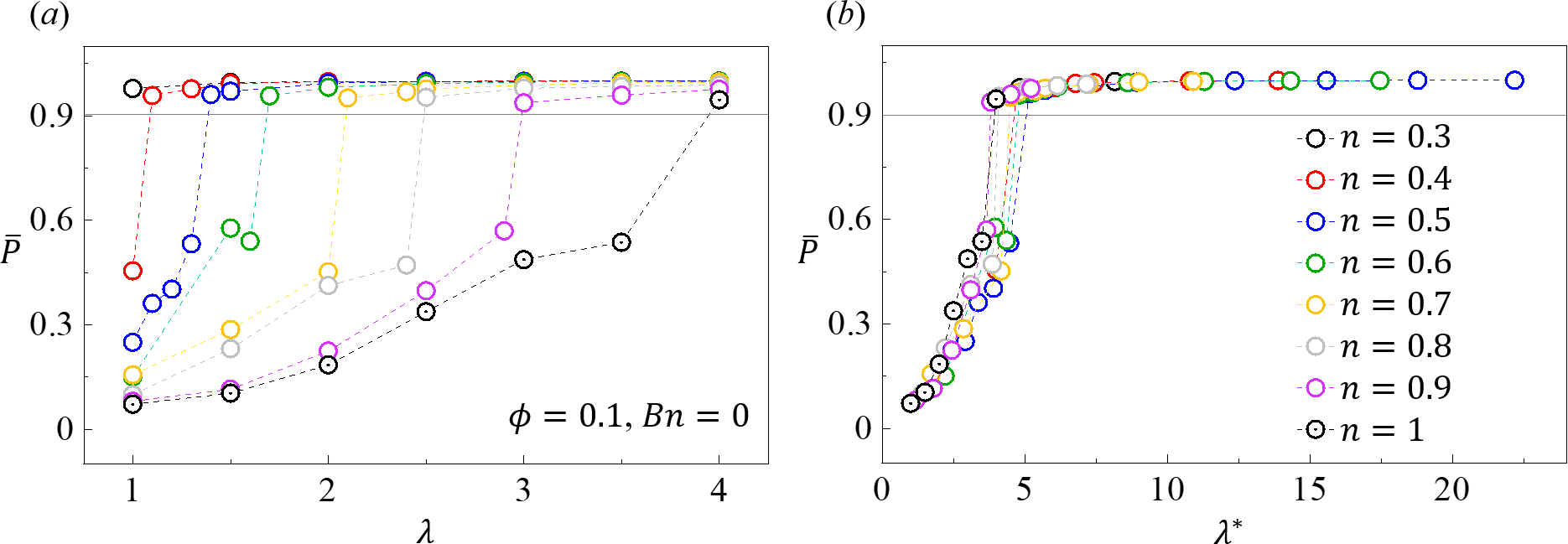}}
\caption{The value of $\bar P$ as a function of ($a$) $\lambda$ and ($b$) $\lambda^*$ for different n ($\phi=0.1$, $Bn=0$).}
\label{figure_curvePvsLamBn}
\end{figure}

Figure \ref{figure_curvePvsLamBn} shows the variation of $\bar P$ as a function of $\lambda$ and $\lambda^*$ for different $n$ ($\phi=0.1$, $Bn=0$). In figure \ref{figure_curvePvsLamBn}($a$), the curves of $\bar P$ are also scattered and the critical $\lambda$ decreases with decreasing $n$. After the rescaling is performed, the curves collapse onto a single curve as a whole. The critical $\lambda^*$ is distributed in a narrow range of $\lambda^* \approx 3.8-5.1$. $\lambda^*$ is more suitable for predicting the critical condition of the FU regime formation for different $n$ than that for different $Bn$.

\section{Conclusions}\label{sec:conclusions}

\noindent In this work, the hydrodynamic coupling of a cilia-mucus system in Herschel-Bulkley flows was numerically investigated using a two-dimensional hydrodynamic model. The mucus flow was predicted based on the lattice-Boltzmann method and the interaction between the cilia and the mucus was handled by an alignment rule. Numerical simulations were performed in a wide range of ciliary density ($\phi$), interaction length ($\lambda$), Bingham number ($Bn$), and flow index ($n$) to highlight the effects of yield stress and shear thinning properties on the mucus flow regime. For the effects of $\phi$ and $\lambda$, a poorly organized (PO) regime, a swirly (S) regime, and a fully unidirectional (FU) regime were identified. The PO regime appears with low $\lambda$ and $\phi$. The S regime appears with low $\lambda$ and high $\phi$. The FU regime appears with high $\lambda$ and high $\phi$. These are determined by the range of influence of the ciliated cells (range of momentum diffusion) and the coordination between different cilia. For the effects of $Bn$ and $n$, the range of influence of the ciliated cells is increased by increasing $Bn$ and decreasing $n$, resulting in the activation of the S and FU regimes at lower $\phi$ and $\lambda$. Mucus viscosity is found to be the dominant parameter affecting the regime formation when varying $Bn$ and $n$. We define an effective interaction length $\lambda^*$ based on the spatially averaged viscosity obtained from the final steady solution instead of the reference viscosity, which is more appropriate than $\lambda$ to represent the range of influence of the ciliated cells. $\lambda^*$ increases with increasing $Bn$ and decreasing $n$, explaining the regime formation upon introduction of Herschel-Bulkley flows. After rescaling, the critical $\lambda^*$ for the appearance of the FU regime are still scattered for different $Bn$, while the critical $\lambda^*$ are distributed in a narrow range for different $n$. $\lambda^*$ is more suitable to predict the critical condition of FU regime formation for different $n$ than that for different $Bn$. Furthermore, the present model is capable of reproducing the solid body rotation observed in experiments, showing a more precise prediction than that of a Newtonian model for the mucus.

\vspace{2ex} \noindent \textbf{Acknowledgements.} Centre de Calcul Intensif d'Aix-Marseille University is acknowledged for granting access to its high performance computing resources. The authors thank Dr. Simon Gsell for his kind assistance on the code.

\vspace{2ex} \noindent \textbf{Funding.} This work was supported by the BonchoClogDrain project (ANR-22-CE30-0045) funded by the French National Research Agency (ANR).

\vspace{2ex} \noindent \textbf{Declaration of interests.} The authors report no conflict of interest.

\bibliographystyle{jfm}
\bibliography{jfm-instructions}

\end{document}